\documentclass[12pt]{article}
\pdfoutput=1

\usepackage{amsmath}
\usepackage{amssymb}
\usepackage{graphicx}
\usepackage{cite}
\usepackage{multirow}
\usepackage{longtable}
\usepackage{lscape}
\usepackage{rotating}
\usepackage{xcolor}
\definecolor{red}{rgb}{1.0, 0, 0}

\allowdisplaybreaks[4] 

\newcommand{\capdef}{}
\newcommand{\mycaption}[2][\capdef]{\renewcommand{\capdef}{#2}%
        \caption[#1]{{\footnotesize #2}}}
\makeatletter
\renewcommand{\fnum@table}{\textbf{\tablename~\thetable}}
\renewcommand{\fnum@figure}{\textbf{\figurename~\thefigure}}
\makeatother

\newcommand{\ie}{{\it i.e.}}

\newcommand{\eg}{{\it e.g.}}

\newcommand{\etc}{{\it etc.}}

\newcommand{\fig}{Fig.}

\newcommand{\Ref}{Ref.}
\newcommand{\Refs}{Refs.}
\newcommand{\Sec}{Sec.}
\newcommand{\Secs}{Secs.}
\newcommand{\App}{Appendix}

\newcommand{\Tab}{Table}

\newcommand{\figu}[1]{\fig~\ref{fig:#1}}

\newcommand{\bi}{\begin{itemize}}
\newcommand{\ei}{\end{itemize}}

\newcommand{\be}{\begin{equation}}
\newcommand{\ee}{\end{equation}}
\newcommand{\bea}{\begin{eqnarray}}
\newcommand{\eea}{\end{eqnarray}}

\newcommand{\ldm}{\Delta m_{31}^2}

\newcommand{\deltacp}{\delta}
\newcommand{\stheta}{\sin^2 2 \theta_{13}}

\catcode`\@=11 
\def\parenbar{\mathpalette\p@renb@r}
\def\p@renb@r#1#2{\vbox{%
  \ifx#1\scriptscriptstyle \dimen@.7em\dimen@ii.2em\else
  \ifx#1\scriptstyle \dimen@.8em\dimen@ii.25em\else
  \dimen@1em\dimen@ii.4em\fi\fi \offinterlineskip
  \ialign{\hfill##\hfill\cr
    \vbox{\hrule width\dimen@ii}\cr
    \noalign{\vskip-.3ex}%
    \hbox to\dimen@{$\mathchar300\hfil\mathchar301$}\cr
    \noalign{\vskip-.3ex}%
    $#1#2$\cr}}}
\catcode`\@=12 

\newcommand{\thk}{{\sf T2HK}}
\newcommand{\lbne}{{\sf LBNE}}
\newcommand{\wbb}{{\sf WBB}}
\newcommand{\bb}{{\sf BB350}}
\newcommand{\lenf}{{\sf NF10}}

\newcommand{\minilbne}{{\sf LBNE$_{\mathsf{mini}}$}}  
\newcommand{\novaplus}{{\sf NO$\nu$A$^+$}} 
\newcommand{\nova}{{\sf NO$\nu$A}}
\newcommand{\tk}{{\sf T2K}}
\newcommand{\lbno}{{\sf LBNO}}
\newcommand{\lenfhs}{{\sf NF5}}  
\newcommand{\bbsb}{{\sf BB+SPL}}
\newcommand{\bblow}{{\sf BB100}}
\newcommand{\spl}{{\sf SPL}}

\newcommand{\db}{{\sf Daya Bay}}
\newcommand{\dc}{{\sf Double Chooz}}
\newcommand{\reno}{{\sf RENO}}
\newcommand{\minos}{{\sf MINOS}}
\newcommand{\ino}{{\sf INO}}
\newcommand{\pingu}{{\sf PINGU}}

\newcommand{\Ne}{$^{18}$Ne }
\newcommand{\He}{$^{6}$He}

\newcommand{\bbeam}{beta-beam}
\newcommand{\bbeams}{beta-beams}

\begin{document}
\begin{titlepage}

\renewcommand{\thefootnote}{\alph{footnote}}

\vspace*{-3.cm}
\begin{flushright}

EURONU-WP6-12-53  \\
FERMILAB-PUB-12-509-T\\
IDS-NF-036\\

\end{flushright}

\renewcommand{\thefootnote}{\fnsymbol{footnote}}
\setcounter{footnote}{-1}

{\begin{center}
{\large\bf
Systematic uncertainties in long-baseline neutrino oscillations for large $\boldsymbol{\theta_{13}}$
} \end{center}}
\renewcommand{\thefootnote}{\alph{footnote}}

\vspace*{.3cm}
{\begin{center}
{\large \today}
\end{center}
}

{\begin{center} {
                \large{\sc
                 Pilar~Coloma\footnote[1]{\makebox[1.cm]{Email:}
                 pcoloma@vt.edu}},
		 \large{\sc
                 Patrick~Huber\footnote[2]{\makebox[1.cm]{Email:}
                 pahuber@vt.edu}}, \\
                 \large{\sc
		 Joachim~Kopp\footnote[3]{\makebox[1.cm]{Email:}
                 jkopp@fnal.gov}, and
                 Walter~Winter\footnote[4]{\makebox[1.cm]{Email:}
                 winter@physik.uni-wuerzburg.de}}
		 }
\end{center}
}
\vspace*{0cm}
{\it
\begin{center}

\footnotemark[1]${}^,$\footnotemark[2]
       Center for Neutrino Physics, Virginia Tech, Blacksburg, VA 24061, USA \\
\footnotemark[3]
       Fermilab, P.O. Box 500, Batavia, IL 60510-0500, USA  and \\
       Max-Planck-Institut f\"ur Kernphysik, PO Box 103980, 69029 Heidelberg, Germany \\
\footnotemark[4]
       Institut f{\"u}r Theoretische Physik und Astrophysik, 
       Universit{\"a}t W{\"u}rzburg, \\
       D-97074 W{\"u}rzburg, Germany

\end{center}}

\vspace*{0.5cm}

{\Large \bf
\begin{center} Abstract \end{center}  }

We study the physics potential of future long-baseline neutrino
oscillation experiments at large $\theta_{13}$, focusing especially on
systematic uncertainties. We discuss superbeams, \bbeams, and
neutrino factories, and for the first time compare these experiments
on an equal footing with respect to systematic errors. We explicitly
simulate near detectors for all experiments, we use the same
implementation of systematic uncertainties for all experiments, and we
fully correlate the uncertainties among detectors, oscillation
channels, and beam polarizations as appropriate.  As our primary
performance indicator, we use the achievable precision in the
measurement of the CP violating phase $\deltacp$. We find that a
neutrino factory is the only instrument that can measure $\deltacp$
with a precision similar to that of its quark sector counterpart.  All
neutrino beams operating at peak energies $\gtrsim 2$~GeV are quite
robust with respect to systematic uncertainties, whereas especially
\bbeams\ and \thk\ suffer from large cross section uncertainties in
the quasi-elastic regime, combined with their inability to measure the
appearance signal cross sections at the near detector. A noteworthy
exception is the combination of a $\gamma=100$ \bbeam\ with an
\spl-based superbeam, in which all relevant cross sections can be
measured in a self-consistent way. This provides a performance, second
only to the neutrino factory. For other superbeam experiments such as
\lbno\ and the setups studied in the context of the \lbne\ reconfiguration
effort, statistics turns out to be the bottleneck.  In almost all
cases, the near detector is not critical to control systematics since
the combined fit of appearance and disappearance data already
constrains the impact of systematics to be small provided that the
three active flavor oscillation framework is valid.

\vspace*{.5cm}

\end{titlepage}

\newpage

\renewcommand{\thefootnote}{\arabic{footnote}}
\setcounter{footnote}{0}

\section{Introduction}
\label{sec:intro}

The story of large $\theta_{13}$ has unfolded in fast succession from
first hints in global
fits~\cite{Fogli:2008jx,Schwetz:2008er,GonzalezGarcia:2010er,Schwetz:2011qt},
via direct indications from \tk~\cite{Abe:2011sj},
\minos~\cite{Adamson:2011qu} and \dc~\cite{Abe:2011fz} to a
discovery by \db~\cite{An:2012eh}, which was soon confirmed by
\reno~\cite{Ahn:2012nd}. 
A recent global fit yields
$\sin^2 \theta_{13}=0.023 \pm 0.0023$~\cite{GonzalezGarcia:2012sz} (see also Refs.~\cite{Fogli:2012ua,Tortola:2012te}, which find very similar values),
where the error bars are entirely
dominated by the reactor measurements. The precision of reactor
experiments on $\theta_{13}$ will continue to improve and
is not expected to be exceeded by beam experiments anytime soon (see,
for instance, \Ref~\cite{Coloma:2012wq}).

The most important open questions in neutrino oscillations, within the context
of three active flavors, are the determination of the neutrino mass hierarchy
($\mathrm{sgn}(\ldm)$) and the measurement of the CP violating phase
$\deltacp$. While there might be already some weak evidence for
$\deltacp \sim \pi$ from global
fits~\cite{Fogli:2012ua,Tortola:2012te}, high confidence level CP
violation (CPV) and mass hierarchy measurements cannot be performed with
existing facilities, such as \db, \reno, \dc, \tk, and
\nova\ in spite of the relatively large value of
$\theta_{13}$~\cite{Huber:2009cw}. In the most aggressive scenario,
\ie, for upgraded proton drivers for both \tk\ and \nova\ and mutually
optimized neutrino-antineutrino running plans, CPV could be
established at $3\sigma$ confidence level only for 25\% of all values of $\deltacp$.
Therefore, a next generation of experiments is mandatory and a decision
towards one of the proposed technologies---superbeam upgrades, a \bbeam\
or a neutrino factory---will soon be needed.

The determination of the mass hierarchy need not necessarily be
performed in long-baseline experiments, given the relatively large
value of $\theta_{13}$. An independent determination of the mass
hierarchy may be provided from the combination of \tk, \nova\ and
\ino~\cite{Blennow:2012gj}, from new proposals such as
\pingu~\cite{Akhmedov:2012ah}, from a reactor experiment with a relatively
long
baseline~\cite{Petcov:2001sy,Choubey:2003qx,Zhan:2009rs},\footnote{This
may be rather challenging from the experimental point of view, see
\Ref~\cite{Qian:2012xh} for instance.} or from the combination of
reactor and long baseline experiments with very high
precision~\cite{Nunokawa:2005nx,deGouvea:2005hk}.  Almost all of the
long-baseline experiments studied in this work would allow for a high
confidence level mass hierarchy discovery because of the sufficient
length of the baselines and the chosen neutrino energies (see, for
instance, Refs.~\cite{Barger:2007jq,Winter:2008cn,Tang:2009wp}).
Those setups with shorter baselines $\lesssim 500$~km,
where it is not possible to determine the mass hierarchy from
the long-baseline data alone, would have a very massive detector. In
these cases, a large sample of atmospheric neutrino events will be
available which in combination with the beam data allows for an
extraction of the mass
hierarchy~\cite{Huber:2005ep,Campagne:2006yx,Abe:2011ts,Barger:2012fx}.
Therefore, we will not focus on this observable in this study.

Regarding $\deltacp$, the main focus in the literature so far has been
on the question whether CPV can be detected, \ie, whether
the CP conserving cases $\deltacp = 0, \pi$ can be excluded.  The discovery of
leptonic CPV would support thermal
leptogenesis~\cite{Fukugita:1986hr}, which could potentially lead to
an explanation of the observed baryon--antibaryon asymmetry of the
Universe---although a direct connection to the CP violating phases in
the high energy theory can only be established in a model-dependent
way.  The CP asymmetry in vacuum is linearly proportional to
$\sin\delta$, and great efforts have been made to optimize neutrino
oscillation facilities for maximal sensitivity to this
term. However, there are good reasons why $\cos \deltacp$ is also
interesting. For example, if the neutrino mass matrix is determined by
a symmetry to have the tri-bimaximal (bi-maximal) form, corrections
originating from the charged lepton mass matrix may lead to the sum
rule~\cite{King:2005bj,Masina:2005hf,Antusch:2005kw} $ \theta_{12}
\simeq 35^\circ \, (45^\circ ) + \theta_{13} \, \cos \deltacp \, .$ It
is obvious that establishing such sum rules, which usually depend on $\cos
\deltacp$, requires the measurement of the $\cos \deltacp$-term. An
ideal long-baseline experiment would therefore have a relatively ``flat''
performance independent of $\deltacp$ and would be able to measure both terms with
similar precision.  The ultimate goal will be the measurement of
$\deltacp$ with a precision comparable to the one achieved in the quark sector.
In order to capture the whole parameter space in
$\deltacp$ for fixed $\theta_{13}$ (or a relatively small range of
$\theta_{13}$), so-called ``CP patterns'' were proposed in
\Ref~\cite{Winter:2003ye,Huber:2004gg} to quantify the achievable
precision as a function of the true $\deltacp$. In
\Ref~\cite{Coloma:2012wq} this dependence was studied in detail for
different types of experiments, and the main factors that affect the
achievable precision were identified. From the results presented in
\Ref~\cite{Coloma:2012wq} it is clear that especially narrow band
beams and setups with short baselines are typically optimized for the
CPV measurement, \ie, a good precision in the measurement of
$\deltacp$ around the particular values $0$ and $\pi$. On the other
hand, more complicated (asymmetric) patterns arise in wide-band beams
or in the presence of matter effects.

The key issue for long-baseline experiments at large $\theta_{13}$ is
systematics. It is well known that especially signal normalization
uncertainties affect neutrino oscillation measurements for large
$\theta_{13}$, see, \eg,
\Refs~\cite{Huber:2007em,Barger:2007jq,Coloma:2011pg}. While in
phenomenological studies near detectors are only in rare cases
explicitly included or discussed, see \eg\ \Ref~\cite{Huber:2007em}
for \thk\ and \Ref~\cite{Tang:2009na} for the neutrino factory, it is
usually assumed that these can be described by an effective systematic
error in the far detector in the range from 1\% to 10\%.  The chosen
values are ``educated guesses'' in the absence of explicit near
detector simulations. This is unsatisfactory given the large impact of
systematic uncertainties at large $\theta_{13}$.
Indeed, it is not even sufficient to use realistic numbers for the
systematic errors, but it is equally important to implement
them in an appropriate way, in particular taking into account correlations
between the errors affecting different oscillation channels,
different parts of the energy spectrum, etc.
For instance, most conventional
simulations assume that systematics are uncorrelated among all
oscillation channels, but fully correlated among all energy bins and
backgrounds. In the real world, cross sections are correlated among
all channels measuring the same final flavor, fluxes among all
channels in the same beam, \etc\ Furthermore, it is known that the
matter density uncertainty affects the measurements for large
$\theta_{13}$ for experiments with long baselines and high energies,
see, \eg, \Refs~\cite{Huber:2002mx,Ohlsson:2003ip}.

In this study, we will explore the effect of systematic errors on the
achievable precision in different experiments, and we will provide a detailed
comparison between different setups under the same assumptions for
the systematics. Our systematics treatment is an extension of the one
used for multi-detector reactor experiment simulations~\cite{Huber:2003pm,
Huber:2004ug, Huber:2006vr}. In particular,
\begin{enumerate}
\item we use a detailed, physics-based and self-consistent systematics
  implementation including correlations, which is
  comparable for all experiments;
\item we explicitly simulate the near detectors, with comparable assumptions
  regarding statistics and geometry for all experiments;
\item we do not only choose particular values for the systematic
  errors, but we also study ranges which span the gamut from
  conservative to  optimistic;
\item we use exactly the same assumptions for cross section and matter
  density uncertainties for all experiments. For the systematic
  uncertainties that depend on the particular type of neutrino beam
  (for instance, flux uncertainties or intrinsic beam backgrounds) we
  consider the same values for all experiments of the same type.
\end{enumerate}
To facilitate the comparison between different facilities, we will use as a
performance indicator the fraction of possible values of $\deltacp$ for which a certain
precision can be obtained, similar to earlier figures showing the CPV
performance. We thus treat the whole parameter space on equal footing, and our
conclusions on the relative sensitivity of different experiments will not
depend on any assumed ``true'' value of $\deltacp$.\footnote{It should be kept
in mind, however, that the experiment which reaches the best overall precision
in $\delta$ may not yield the best CPV discovery potential (and {\it vice
versa}), since the latter depends on the achievable precision around the
specific values $\delta=0,\pi$.}

For the experiment definitions and simulations, we have modified the AEDL
language (Abstract Experiment Definition Language) of the GLoBES
software~\cite{Huber:2004ka,Huber:2007ji}, which allows now for a flexible
systematics implementation entirely in AEDL (without the need to write C
code).\footnote{This is one of the key modifications of the software which is
expected to be included in the GLoBES~4.0 release.}

The paper is organized as follows.  In \Sec~\ref{sec:simulation} the
experimental setups are described, as well as the assumptions for the
oscillation parameters, the treatment of systematics, the values
chosen for the systematic errors and the definition of our performance
indicator. In \Sec~\ref{sec:systematics}, we compare the
results obtained with the new systematics implementation (using
explicit near detector simulations and including correlations) to those
obtained with the old implementation (using an effective description of the
errors in the far detector).  More details on the
simulation of the various experiments can be found in
appendix~\ref{sec:sim}, and the details on our statistical methods
can be found in appendix~\ref{sec:sysdet}.  We also illustrate for
which experiments robust predictions can be made and for which ones
more external information is needed. A comparison of the performance
of all setups is presented in \Sec~\ref{sec:performance}, where also
the dependence on exposure is discussed. In \Sec~\ref{sec:impacts} we
identify for each experiment the relevant performance bottlenecks
and provide guidance on which quantities should be optimized in each case.
Finally, we summarize and conclude in \Sec~\ref{sec:summary}.

\section{Simulation techniques and systematics treatment}
\label{sec:simulation}

\subsection{Experimental setups}
\label{sec:setups}

\Tab~\ref{tab:setups} summarizes the main features of the setups
studied in this work. We have chosen four representative
\textit{benchmark} setups for long baseline neutrino oscillation
experiments:
\begin{description}
\item[Beta-beam:] a high-$\gamma$ ($\gamma=350$)
  \bbeam~\cite{BurguetCastell:2003vv,BurguetCastell:2005pa} has
  been chosen, since it provides a very good CPV discovery
  potential, even comparable to the one obtained at a neutrino factory, see 
   {\it e.g.}\ \Ref~\cite{Bernabeu:2010rz}. The relatively long
  baseline ($L=650$ km) is enough to guarantee a measurement of the
  mass hierarchy given the large value of
  $\theta_{13}$ (see for instance \Ref~\cite{Winter:2008cn}). The beam is aimed at a 500~kton
  Water \v{C}erenkov (WC) detector. This setup will be referred to as
  \bb .
\item[Neutrino factory:] we consider a low energy version of the
  neutrino factory, with a parent muon energy of 10~GeV and a
  Magnetized Iron Neutrino Detector (MIND) detector placed at a
  baseline of 2\,000~km~\cite{Agarwalla:2010hk}. This is the setup
  currently under consideration within the International Design Study
  for a Neutrino Factory (IDS-NF)~\cite{ids}. It will be
  referred to as \lenf\ hereafter.
\item[Off-axis conventional neutrino beam:] here we follow
  the \thk\ proposal given its high relevance in the literature and that a
  Letter of Intent (LoI) has already been submitted~\cite{Abe:2011ts}.  The
  experiment uses a WC detector with a fiducial mass of 560~kton, placed at a
  distance of 295~km from the source.
\item[On-axis conventional neutrino beam:] we study a setup
  with a relatively high-energy flux (taken from \Ref~\cite{Longhin:2010zz})
  and with a 100~kt Liquid Argon (LAr) detector at 2\,300~km from the source.
  This corresponds to one of the configurations under consideration within
  LAGUNA~\cite{laguna} and LAGUNA-LBNO~\cite{LBNOeoi}. We have checked that the
  Fermilab-to-DUSEL Long Baseline Neutrino Experiment (\lbne), with a 34~kt LAr
  detector at a baseline of 1\,290~km~\cite{Akiri:2011dv}, has a very similar
  performance.  We will therefore refer to this setup \wbb\ in the rest of
  this paper since the conclusions extracted from its performance would be
  generally applicable to both \lbno\ and \lbne.\footnote{We find a slightly worse
  performance for the \lbne\ setup, though. }
\end{description}
%

\begin{table}[t!]
\begin{center}{
\renewcommand{\arraystretch}{1.5}
\begin{tabular}{l|lllclcccc}
& Setup & $E^\textrm{peak}_\nu$ & $L$ & OA & Detector & kt & MW & Decays/yr & ($t_\nu$,$t_{\bar\nu}$)  \\
\hline
\multirow{4}{*}{\begin{sideways} Benchmark\phantom{l}  \end{sideways}} 
& \bb\ &  1.2  & 650  & -- & WC   & 500 & --   & 1.1(2.8)$\times10^{18}$ & (5,5) \cr \cline{2-10}
& \lenf\ & 5.0  & 2\,000 & -- & MIND & 100 & --   & 7$\times10^{20}$ & (10,10)  \cr \cline{2-10}
& \wbb\  & 4.5 & 2\,300 & -- & LAr  & 100 & 0.8  & -- & (5,5) \cr \cline{2-10}
& \thk\  & 0.6 & 295  & $2.5^\circ$  & WC   & 560 & 1.66 & -- & (1.5,3.5)  \\ \hline

\multirow{4}{*}{\begin{sideways} Alternative\phantom{lalaa}  \end{sideways}}
& \bblow  & \multirow{2}{*}{0.3}  & \multirow{2}{*}{130} & -- & \multirow{2}{*}{WC} & \multirow{2}{*}{500} & --  & 1.1(2.8)$\times10^{18}$ & (5,5) \cr
& + \spl  &  & & -- & & & 4 & -- & (2,8) \\ \cline{2-10}
& \lenfhs & 2.5  & 1\,290 & -- & MIND & 100 & --  & 7$\times10^{20}$ & (10,10) \\ \cline{2-10}
& \minilbne & 4.0 & 1\,290 & -- & LAr  & 10  & 0.7 & -- & (5,5) \\ \cline{2-10}
& \novaplus & 2.0 & 810  & $0.8^\circ$ & LAr  & 30  & 0.7 & -- & (5,5)  \\ \hline
\multirow{2}{*}{\begin{sideways} 2020\phantom{l}  \end{sideways}}
& \tk     & 0.6 & 295 & $2.5^\circ$ & WC & 22.5  & 0.75 & -- & (5,5)  \cr
& \nova & 2.0 & 810  & $0.8^\circ$ & TASD  & 15  & 0.7 & -- & (4,4)  \\ \hline
\end{tabular}}
\end{center}
\mycaption{\label{tab:setups} Main features of the setups considered in this
work. From left to right, the columns list the names of the setups, the approximate
peak energy of the neutrino flux, the baseline, off-axis angle, detector technology,
fiducial detector mass, beam power (for the conventional and superbeams) or
useful parent decays per year (ion decays for the \bbeams\ and muon
decays for the neutrino factories), and the running time in years for
each polarity. Note, that the neutrino and antineutrino running is
simultaneous for the neutrino factory setups \lenf\ and \lenfhs\ (the $\mu^+$ and $\mu^-$ circulate in different directions within the ring). For \bbeams\
the number of useful ion decays is different for the two polarities, so we
quote the number of useful \Ne (\He) decays per year separately. For details on our
simulations, see Appendix~\ref{sec:sim}.  }
\end{table}

In addition to these setups, we will also considered four alternative setups
with high relevance in the literature. In particular, we will discuss two out of
the three options considered during the \lbne\ reconfiguration process: a new
Fermilab-based beam line aimed at a 10~kt LAr surface detector placed at
Homestake (\minilbne), and the existing NuMI beam with a new 30~kt LAr surface
detector placed at the Ash River site (\novaplus).\footnote{The third option
considered within the \lbne\ reconfiguration process consists of a 15~kt LAr
underground detector placed at the \minos\ site in Soudan. However, 
we have checked that the performance of this setup is much inferior to that of all other setups
discussed here, and therefore we do not consider it any further.} 
Moreover, we consider a lower energy version of the
neutrino factory (\lenfhs), with a muon energy of 5~GeV and a baseline of
1\,300~km.  (The detector technology in this case is still a MIND to make a
direct comparison to the \lenf\ setup easier.) The fourth alternative setup
discussed in this paper is combination of a low-$\gamma$ ($\gamma=100$)
\bbeam\ with the \spl~\cite{Campagne:2006yx}, labeled as \bbsb. We use this
setup to study whether a combination of different channels (in this case,
CPT conjugates) can reduce the impact of systematic errors. More details about 
each setup are given in Appendix~\ref{sec:sim}.

Finally, sometimes we will compare the results for the setups listed in
\Tab~\ref{tab:setups} to the results that would be obtained by 2020 from the
combination of present facilities, that is, \tk, \nova\ and reactors. In order to
do so, we assume that \tk\ and \nova\ will have run for 5 and 4 years per polarity
by that date, respectively, and that the precision on the measurement of
$\theta_{13}$ will be dominated by the systematic error reachable at \db.
In order to simulate this combination we have followed
\Ref~\cite{Huber:2009cw}. 

It was noted in \Ref~\cite{Coloma:2012wq} that the achievable precision in
$\delta$ around $\delta=\pm 90^\circ$ is compromised in \bbeam\ setups
because, unlike superbeams or neutrino factories, they cannot obtain a precise
measurement for the atmospheric parameters through the $\nu_\mu$ disappearance
channels (see also \Ref~\cite{Donini:2005rn}). Therefore, we have also combined
the data obtained at \bb\ with the disappearance data expected from \tk.  Note
that this is not necessary for \bbsb\, since in this case the \spl\ beam would
already provide a better measurement of the atmospheric parameters.

For all setups, we assume the near detector to be sufficiently far away from
the neutrino production region to have the same geometric acceptance as the far
detector. (A dedicated study would be needed to determine the appropriate
distance for each setup.) This ensures that the ratio of near and far detector
event rates is independent of energy.  Apart from the baseline and mass, the
near and far detectors are considered to be identical, \ie, they share the same
energy resolutions, efficiencies,\footnote{The treatment of NC backgrounds takes
into account possible differences, though, see Appendix~\ref{sec:sim} for
details.} energy binnings, \etc

\subsection{Input values for the oscillation parameters }

The following input values for the oscillation parameters, in
agreement with the allowed ranges at $1\sigma$ from global
fits~\cite{GonzalezGarcia:2012sz,Fogli:2012ua,Tortola:2012te}, are used for all simulations
in this paper:
\begin{eqnarray}
 &\Delta m_{21}^2 = 7.64\times10^{-5} \textrm{eV}^2, \quad
 \Delta m_{31}^2 = 2.45\times10^{-3}  \textrm{eV}^2 ; \nonumber \\
 &\theta_{21} = 34.2^\circ ,\quad
 \theta_{23} = 45^\circ , \quad
 \theta_{13} = 9.2^\circ . \nonumber  
\end{eqnarray}
Unless otherwise stated, a normal mass hierarchy is assumed.  In our fits, we
include Gaussian priors with a $1\sigma$ width of 3\% for the solar parameters,
8\% for $\theta_{23}$ and 4\% for $\Delta m_{31}^2$.  No external priors were
used for $\theta_{13}$ and $\delta$. We have checked that adding a prior on
$\theta_{13}$ corresponding to the expected precision of the final \db\
measurement~\cite{An:2012eh} does not affect our results, except for a very
mild improvement in the measurement of $\delta$ around $ \pm 90^\circ$ for some
facilities where the intrinsic degeneracy is still present.

We compute the neutrino scattering cross sections in the target as a
function of neutrino energy using GENIE version
2.6.0~\cite{Andreopoulos:2009rq}. We split the cross sections into
Neutral Current (NC) and Charged Current (CC) contributions, and we further
subdivide the latter into the three regimes Quasi-Elastic scattering (QE),
RESonance production (RES), and Deep-Inelastic Scattering (DIS).  The cross
sections per nucleon vary by $\mathcal{O}(10\%)$ between targets with
different proton-to-neutron ratios and nuclear masses, but for easier
comparability among experiments, we use cross sections for $^{28}$Si,
which is an intermediate mass, isoscalar nucleus, throughout.

As a further simplification, no sign degeneracies have been considered in this
analysis.  This motivated by the fact that, thanks to the relatively large
value of $\theta_{13}$, almost all experiments presented in the comparison
would most likely be able to measure the mass hierarchy, either from matter
effects at long baselines (\lenf, \lenfhs, \wbb, \minilbne, \bb) or from the
analysis of atmospheric data at very massive WC detectors (\thk , \bb, \bbsb).
The only exception to this could perhaps be \novaplus, due to its relatively
short baseline (735~km) and limited detector mass. Nevertheless, as already
stated in \Sec~\ref{sec:intro}, an independent determination of the mass
hierarchy may be provided by other means anyway.

\subsection{Systematic errors and their implementation}
\label{sec:systreat}

\begin{table}[t]
  \begin{center}
    \begin{tabular}{p{4.3cm}|rrr|rrr|rrr}
      \hline
       & \multicolumn{3}{c|}{SB } & \multicolumn{3}{c|}{BB} & \multicolumn{3}{c}{NF } \\
      Systematics & Opt. & Def. & Cons. & Opt. & Def. & Cons. & Opt. & Def. & Cons. \\
      \hline
      Fiducial volume ND  & 0.2\% & 0.5\% & 1\% & 0.2\% & 0.5\% &  1\% & 0.2\% & 0.5\% & 1\% \\
      Fiducial volume FD & 1\% & 2.5\% &  5\% & 1\% & 2.5\% & 5\% & 1\% & 2.5\% & 5\%  \\
      (incl. near-far extrap.) & & & & & & & & &  \\
      Flux error signal $\nu$ & 5\% & 7.5\% & 10\% & 1\% & 2\% & 2.5\% &  0.1\% & 0.5\% & 1\%  \\
      Flux error background $\nu$  & 10\% & 15\% & 20\% & \multicolumn{3}{c|}{correlated} & \multicolumn{3}{c}{correlated} \\
      Flux error signal $\bar\nu$ & 10\% & 15\%  & 20\% & 1\% & 2\%  & 2.5\% &  0.1\% & 0.5\% & 1\%  \\
      Flux error background $\bar\nu$  & 20\% & 30\% &  40\% & \multicolumn{3}{c|}{correlated} & \multicolumn{3}{c}{correlated} \\
      Background uncertainty & 5\% & 7.5\% & 10\% & 5\% & 7.5\% & 10\% & 10\% & 15\% & 20\% \\ 
      \hline
      Cross secs $\times$ eff. QE$^\dagger$ & 10\% & 15\% & 20\% & 10\% & 15\% & 20\% & 10\% & 15\% & 20\%  \\
      Cross secs $\times$ eff. RES$^\dagger$ & 10\% & 15\% & 20\% & 10\% & 15\% & 20\% & 10\% & 15\% & 20\%  \\
      Cross secs $\times$ eff. DIS$^\dagger$ & 5\% & 7.5\% & 10\% & 5\% & 7.5\% & 10\% & 5\% & 7.5\% & 10\%  \\
      \addtocounter{footnote}{1}%
      Effec.~ratio $\nu_e$/$\nu_\mu$  QE$^\star$ & 3.5\% & 11\% & --\phantom{la} & 3.5\% & 11\% & --\phantom{la} & --\phantom{la} & --\phantom{la} & --\phantom{la} \\
      Effec.~ratio $\nu_e$/$\nu_\mu$ RES$^\star$ & 2.7\% & 5.4\% & --\phantom{la} & 2.7\% & 5.4\% & --\phantom{la} & --\phantom{la} & --\phantom{la} & --\phantom{la}  \\
      Effec.~ratio $\nu_e$/$\nu_\mu$ DIS$^\star$ & 2.5\% & 5.1\% & --\phantom{la} &  2.5\% & 5.1\% & --\phantom{la} & --\phantom{la} & --\phantom{la} & --\phantom{la} \\
      \hline
      Matter density &1\% & 2\% & 5\%&1\% & 2\% & 5\%&1\% & 2\% & 5\%\\
      \hline
    \end{tabular}
  \end{center}

  \mycaption{\label{tab:values} The systematic errors considered in our
  analysis for superbeams (SB), \bbeams\ (BB), and neutrino factories (NF),
  respectively. Numerical values are shown for optimistic, default, and
  conservative assumptions.  All numbers are based on external input and do not
  yet include any information from the near detector.  Note that the background
  uncertainties listed here affect only detector-related backgrounds (NC
  events, charge or flavor misidentification), whereas the uncertainties
  related to the intrinsic beam background for superbeams are treated as flux
  errors.
  \newline
  $^\dagger$Despite showing only a single entry here, we use independent
  nuisance parameters for the $\nu_\mu$, $\nu_e$, $\bar\nu_\mu$, and
  $\bar\nu_e$ cross sections.  \newline
  $^\star$Despite showing only a single entry here, we use independent nuisance
  parameters for the $\nu_\mu/\nu_e$, and $\bar\nu_\mu/\bar\nu_e$ cross section
  ratios. The uncertainty due to different detection efficiencies for the
  different lepton flavors has also been added in quadrature, and therefore only 
  the error on the effective ratio is shown. Blank spaces indicate the cases 
  when $\nu_e$ and $\nu_\mu$ cross sections are allowed to vary in a completely independent way. }
\end{table}

In this study, we treat systematic uncertainties for all experiments
in the same framework. We implement beam flux uncertainties,
fiducial mass uncertainties, cross section uncertainties, and the
matter density uncertainty in the same way for all setups, whereas
background uncertainties can only be the same within each class of
experiments---superbeams, \bbeams, and neutrino
factories. For each systematic error, we consider default, optimistic
and conservative values, see \Tab~\ref{tab:values}. Note that the
error estimates given in this Table do \emph{not} include the impact of the
near detectors, which instead are explicitly simulated in our study. Note also
that we do not claim that our default values will actually be exactly realized
by any of the experiments, but the range from optimistic to conservative is
likely to delimit the actual value.

In the following, we describe the types of the systematic errors and
their physical underpinning. (More details are given in \App~\ref{sec:sysdet}.)

\paragraph{Flux errors.} We assume the flux errors to be uncorrelated
among different beam polarities, but fully correlated between the near
and far detectors and between all oscillation channels in the same
beam. For the superbeams, we include an additional flux error for each
of the intrinsic $\nu_e$, $\bar\nu_e$ and wrong-sign $\nu_\mu$
components in the beam. We simulate $\nu_e + \bar\nu_e$ events in both
near and far detectors, so that the former effectively measures the
intrinsic beam backgrounds and moreover provides some information on
the $\nu_e$, $\bar\nu_e$ cross sections.  Note that the precision of
these measurements depends on the size of the near detector because
the event rate from the intrinsic beam contamination is much smaller
than the one from the muon neutrinos in the beam. We choose the
magnitude of the flux errors for superbeams based on what has been
achieved in $\pi$-decay based neutrino beams like the \minos\
experiment~\cite{Adamson:2009ju}.\footnote{Even though our flux
  uncertainties may be optimistic compared to current estimates, their
  effect on the results is not very relevant, as it will be
  demonstrated in Sec.~\ref{sec:impacts}.} For the neutrino factory,
the chosen flux errors correspond to the design values of the
IDS-NF~\cite{NF:2011aa}. For the $\beta$-beams, similar flux errors should be
attainable in principle because for both the neutrino factory and the beta
beam, the parent decays have simple and well understood kinematics, and careful
monitoring of the beam momentum distribution in the storage ring should be
possible.

\paragraph{Cross section errors.} We assume the external
knowledge on cross sections to be universal for all experiments. We
introduce separate systematic uncertainties for the QE, RES and DIS regimes, so
that we have a set of three cross section errors for each of the four
relevant neutrino species $\nu_\mu$, $\bar\nu_\mu$, $\nu_e$, and
$\bar\nu_e$.  By treating the QE, RES and DIS cross sections as
independent, we effectively introduce an uncertainty on the shape of
the neutrino event spectrum.\footnote{Possible shape uncertainties for the cross sections within 
each regime are not considered for simplicity. It also should be pointed out that nuclear 
effects do affect neutrino energy reconstruction~\cite{Lalakulich:2012hs} at low energies. Dedicated migration matrices are needed to study the impact of this effect, though. } In practice, the \emph{ratios} of the
$\parenbar{\nu}_e$ and $\parenbar{\nu}_\mu$ cross sections are known
with greater precision than their absolute values, especially at high
energy where lepton mass effects on the kinematics, effective nuclear
form factors, \etc\ are less important~\cite{Day:2012gb}.
Therefore,
we altogether will have 12 nuisance parameters for the cross sections:
three regimes (QE, RES and DIS) times two flavors ($\nu_e,\nu_\mu$) times two polarities ($\nu$, $\bar\nu$).
In addition we will use six nuisance
parameters for the QE, RES and DIS cross section ratios $\nu_e/\nu_\mu$
and $\bar\nu_e/\bar\nu_\mu$ whenever the final flavor cross section cannot be measured 
at the near detector (\ie, for BB and SB). 
However, in the cases where the constraint on the cross
section ratio is weaker than the corresponding constraint on the two
cross sections, we \emph{omit} the ratio altogether from the $\chi^2$,
and apply the cross section uncertainty directly to the different
flavors (conservative cases in Table~\ref{tab:values}).

All cross section uncertainties are fully correlated among all
channels measuring the same final neutrino flavor. For 
simplicity, we use the same error ranges for both neutrinos and
antineutrinos as well as for electron and muon flavors (but the
errors are still fully independent and uncorrelated between different
flavors and/or polarities).  As a result, the errors for $\nu_\mu$ 
cross sections are generally slightly overestimated while the errors
for $\bar\nu_e$ are underestimated. However, as we will
illustrate later, only the cross section \emph{ratios} have a sizable
impact on the experimental sensitivity, and therefore this
simplification does not affect our results. 

In principle one might assume that, given lepton universality, the
flavor ratios for cross sections should be entirely determined by
kinematics and that, sufficiently far away from thresholds such as the
electron or muon mass, the resulting differences between cross
sections for different flavor should be small. However, there is a
multitude of effects that depend on the lepton mass and kinematics,
in particular when the non-pointlike nature of the nuclear target is
taken into account. Indeed, the cross sections depend on
a set of form factors, all of which have, at least in
principle, some dependence on the momentum transfer $Q^2$ and are thus are quite sensitive to the
details of the lepton momentum distribution. For instance, for
QE events a recent study~\cite{Day:2012gb} addresses these issues in
detail and does indeed confirm differences as high as 10\% for QE reactions at
energies below 1~GeV.  What is more,
the uncertainties on these differences can be of the
order of the difference itself. These uncertainties arise from possible
uncertainties related to pseudo-scalar form factors and their
$Q^2$-dependence.\footnote{This $Q^2$-dependence is usually
  constrained by the Goldberger-Treimann relation, but experimental
  tests of the PCAC hypothesis, which is the foundation for
  Goldberger-Treimann relation, still have sizable uncertainties.}
Another source of uncertainty is the possible presence of second class
currents.  Overall, our assumptions on the uncertainty of the flavor
ratio of the QE cross sections in this paper are based on \Ref~\cite{Day:2012gb}.
There, it is shown that the uncertainty is strongly energy-dependent.
We do not take this energy dependence into account, but our default
value for the uncertainty (10\%) corresponds to the regime
where the QE cross section peaks (0.5~GeV), and we have
chosen conservative (30\%) and optimistic (2.5\%) values taking into
account that it varies over energy and it is difficult to assign a
specific (theoretical) number. For the other two reaction channels, we have followed the discussion in \Ref~\cite{Huber:2007em}, and 
our default values for the cross section ratios have been set at the 
$2\%$ and $1\%$ for the RES and DIS regimes, respectively.

\paragraph{Efficiency errors.} For simplicity,
efficiency errors are not treated as independent, but absorbed into the cross
section uncertainties. Only the product of the cross section and
efficiency for a given flavor and event type appears in observables,
and thus, in our analysis, only the combined effect has been considered.
Obviously, the physics governing the uncertainties on cross sections
and efficiencies is completely different, but their
effects can still be added in quadrature to obtain the error on the product of cross
section and efficiency. This will be referred to as ``effective cross
section'' from here on. We neglect
the uncertainty on the \emph{absolute} efficiencies because the uncertainty
on the absolute cross section is typically much larger. Nevertheless, we do
include a 5\% uncertainty on the effective cross section ratios listed in \Tab~\ref{tab:values}
due to the different detection efficiencies for different lepton flavors. Such uncertainty is 
added in quadrature to
the uncertainty on the cross section ratio.  We do this independently
for neutrinos and antineutrinos and for the QE, RES and DIS regimes. 
As for the optimistic and conservative cases, the corresponding errors have been taken 
to be a factor of two smaller and larger than the default value, respectively.
Note that the particular value chosen here (5\%), which in the following will be used for all experiments in order to
treat them on equal footing, should be considered a rough
estimate -- in reality, the efficiency error can vary widely, depending
for instance on the detector technology, reaction channel, \etc\  We
assume that the near and far detectors are sufficiently similar for
the efficiency errors to be correlated between them. A residual
near--far difference can be at least partially absorbed into the
fiducial mass errors. 

\paragraph{Near--far extrapolation errors.} Using near detector data to
reliably predict the neutrino spectrum at the far detector is quite
challenging in practice, and the extrapolation is affected by a number
of systematic uncertainties, for instance due to different geometric
acceptance, different detector design, \etc\ Some of these
uncertainties moreover have a strong energy dependence, which can,
however, be reduced by carefully designing the experiment; for
instance, the near detector should not be too close to the neutrino
production volume to make its geometric acceptance as similar as
possible to that of the far detector. Here, we assume for simplicity
that the near--far extrapolation errors are energy and flavor
independent, and can therefore be treated as an effective fiducial
mass error, uncorrelated between the two detectors.  We conservatively
assume this error to be fairly large (2.5\% in our default scenario,
included in the fiducial volume error for the far detector), but it will be shown 
in \Sec~\ref{sec:impacts} that it does not affect the
experimental sensitivity significantly.  The same
assumptions regarding systematic uncertainties have again been made
for all experiments in this case.

\paragraph{Background uncertainties.} In addition to the uncertainties
on the intrinsic beam backgrounds discussed above, we also include
uncertainties on detector-related backgrounds such as NC
events and charge or flavor misidentification. These errors are
typically uncorrelated among channels and detectors. It is
well known, though, that for large $\theta_{13}$ the overall impact of
background uncertainties is small.

\paragraph{Matter density uncertainty.} For relatively short baselines
that traverse only the Earth's mantle, it is a good approximation to
assume the matter density to be constant along the baseline.  We compute
the constant effective density separately for each experiment based on
the density profile along the baseline derived from the Preliminary
Reference Earth Model (PREM)~\cite{Dziewonski:1981xy}.  The
mantle density can be determined from seismic wave measurements
with an uncertainty of typically about 5\% or below. For specific
trajectories, values of about 2\% have been
achieved~\cite{Kozlovskaya:2003kk}, so we use this number as our
default. In the conservative case, we use 5\%, whereas in the
optimistic scenario we assume an uncertainty of only 1\%, which may be
achievable with a dedicated geophysical campaign.

\subsection{Performance indicators}
\label{sec:indicator}

\begin{figure}
\begin{center}
\includegraphics[width=\textwidth]{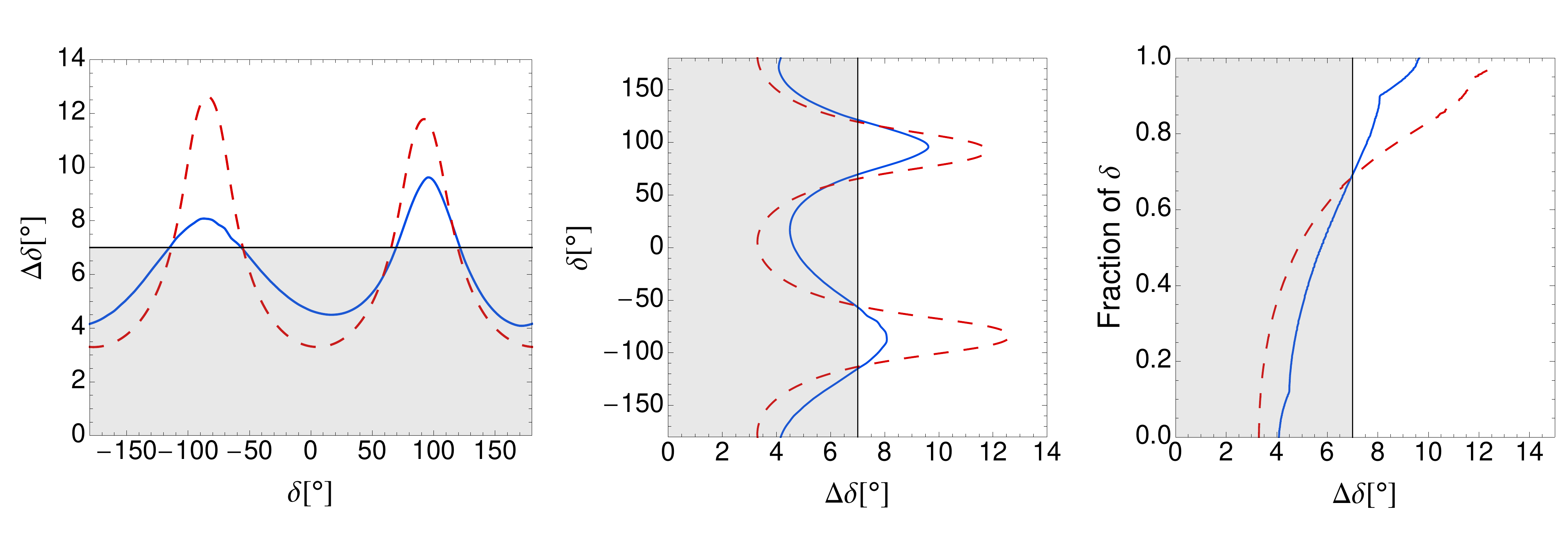}
\end{center}
\mycaption{\label{fig:cpfraction} Left panel: Error on $\deltacp$ ($1
  \sigma$) as a function of the true $\deltacp$ for two sample experiment
  (solid and dashed curves). Middle panel: Left plot mirrored
  at the diagonal. Right panel: Fraction of true $\deltacp$ as a
  function of the precision of $\Delta \delta$ for these two
  experiments. The vertical axis shows for what fraction of the
  $\deltacp$ values a certain precision can be obtained. The lines
  show where the curves intersect.}
\end{figure}

One of the challenges in comparing future experiments based on their achievable
precision in $\delta$, which we will call $\Delta \delta$, is that this
precision strongly depends on $\deltacp$ itself.  In
\Refs~\cite{Winter:2003ye,Huber:2004gg,Coloma:2012wq}, it has
therefore been proposed to show the performance as a function of the true
$\deltacp$, as illustrated in the left panel of \figu{cpfraction}
for two illustrative experiments.
While these two experiments exhibit the same qualitative dependence on
$\deltacp$, the sensitivity of the experiment corresponding to the solid curve
varies less with $\deltacp$. This can be due to a
number of reasons, like the baseline or the width of the energy spectrum (for
a detailed discussion see \Ref~\cite{Coloma:2012wq}), but it is obvious
that both experiments are optimized for their sensitivity to CPV 
since the error is smallest for $\delta=0$ and $\pi$.
Depending on the chosen value of the CP phase one finds that either
the experiment represented by the dashed line or the one corresponding
to the solid line is more precise. If matter effects are strong,
or if the numbers of neutrino and antineutrino events is very different,
the pattern can be shifted in $\delta$, complicating the situation
further~\cite{Coloma:2012wq}.

The new paradigm in neutrino oscillations for large $\theta_{13}$ is
precision, and therefore different experiments should be compared
based on how well they perform on this task within the remaining
parameter space of interest---which is mainly the unknown phase
$\deltacp$. Since it is difficult to compare in a quantitative way the
different experiments using a plot like the left hand panel of
\figu{cpfraction}, we mirror the plot at the diagonal, as shown in the
middle panel, and stack the values of true $\deltacp$ for which a
certain precision can be obtained. The result of this procedure is the
``fraction of $\deltacp$'' (sometimes also called ``CP fraction'') for which $\Delta\delta$ is smaller than a
given number, as illustrated in the right panel of \figu{cpfraction}.
The same approach was previously used to quantify the discovery
potentials for CPV, mass hierarchy, and $\theta_{13}$ by
showing the fraction of true values of $\deltacp$ for which a certain
observable could be measured as a function of the true $\stheta$.
With the fraction of $\deltacp$ on the vertical axis in
\figu{cpfraction} (right), the comparison no longer depends on
relative phase shifts between the two curves from the left panel, it
quantifies the performance assuming that all values of $\deltacp$ are
equally likely and equally important. The disadvantage is that one
cannot read off from the plot at which value of $\deltacp$ the
performance of an experiment peaks. Note, however, that the discussed
setups are typically optimized for CPV, which means that the
optimal performance is in most cases achieved close to $\deltacp = 0$
or $\pi$~\cite{Coloma:2012wq}.

A somewhat more subtle technical point, elaborated on in
Refs.~\cite{Winter:2003ye,Huber:2004gg}, is the fact that the absolute
performance can depend in a nontrivial way on the chosen confidence level. For
instance, $\Delta\chi^2$ may behave in a highly non-Gaussian way far from the
best fit point, in particular if the mass hierarchy degeneracy cannot be
resolved.  Here we assume that we are in the Gaussian limit, and sign degeneracies 
have not been considered. However, we note that the fraction of $\deltacp$
values for which a certain sensitivity is reached is also useful in the
non-Gaussian case.

In the following sections, we will illustrate our results using plots similar
to \figu{cpfraction}, right panel, as well as the standard plots showing the
CPV discovery potential.

\section{Impact of systematics}
\label{sec:systematics}

\begin{figure}
\begin{center}
\includegraphics[width=\textwidth]{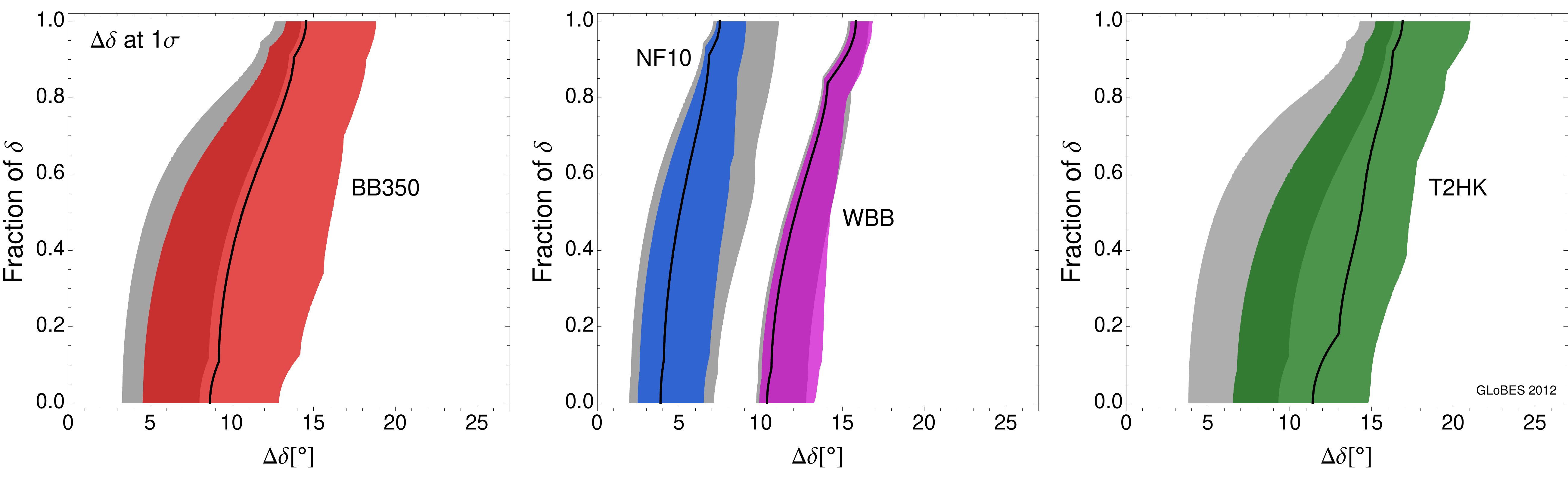} \\
\includegraphics[width=\textwidth]{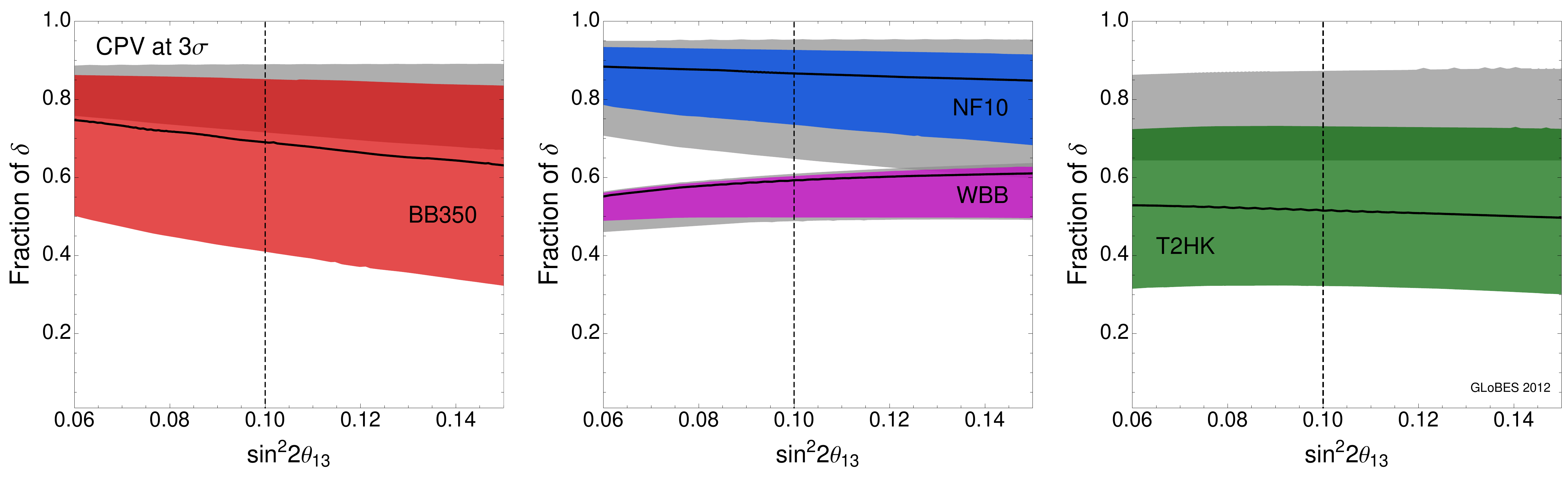}
\end{center}
\mycaption{\label{fig:bands} Comparison of experimental sensitivities predicted with
  the old systematics treatment using effective uncertainties in the far
  detector (light gray shadings) and the new treatment which includes near and
  far detectors and properly accounts for correlations of uncertainties between
  different channels, detectors, \etc\ (dark colored shadings).  The upper row
  shows the fraction of $\deltacp$ values for which $\deltacp$ can be measured
  with a given precision (at $1\sigma$, for $\stheta=0.1$), the lower row shows
  the fraction of $\deltacp$ for which a $3\sigma$ discovery of CPV is possible
  as a function of the true $\stheta$. Different experiments are shown in
  different panels, as indicated in the legends. For the old systematics
  implementation, the effective errors are varied between no systematics and
  5\% normalization error for the signal (10\% for the background), whereas for
  the new implementation the ranges between the optimistic and conservative
  cases from \Tab~\ref{tab:values} are considered. The default values are shown
  as black curves. A true normal hierarchy has been assumed, and no sign
  degeneracies have been considered.  The vertical dotted lines in the lower
  panels correspond to $\stheta=0.1$, which is the true value chosen for
  $\theta_{13}$ in the upper panels.}
\end{figure}

To begin our analysis of systematic uncertainties, we compare in \figu{bands}
the results obtained with our new implementation of systematic errors to the
ones obtained with the simpler, but widely used implementation, in which
systematic errors are not correlated between different oscillation channels and
near detectors are not explicitly simulated. Instead, their effect is
parameterized in terms of effective uncertainties in the far detector, which
are typically chosen between zero (statistical errors only) and $\mathcal{O}
(1-10)\%$.  In our new implementation, the near detectors are explicitly
simulated, the systematic errors are correlated in a physical way, and they
systematic are varied between the optimistic and conservative values from
\Tab~\ref{tab:values}.  Here we focus on the benchmark setups from
\Tab~\ref{tab:setups}, which give an idea on the ultimate performance that each type of beam could reach.
 
The upper row of plots in \figu{bands} shows the fraction of possible values of
$\deltacp$ for which a certain precision $\Delta\delta$ can be achieved (at
$1\sigma$, for $\stheta=0.1$), while the lower row shows the fraction of
possible values of $\deltacp$ for which CPV can be established at $3\sigma$, as
a function of $\stheta$.  The light gray bands show how the performance of each
experiment would vary between the statistics-only (no systematics) limit and
the case where each signal (background) channel in the far detector is assigned
an uncorrelated 5\% (10\%) error.  The matter density uncertainty is also
included in this case: the right/lower edges of the gray bands has been
computed assuming a 5\% matter density uncertainty, whereas for the left/upper
edges the matter density has been kept fixed.  The colored (dark) bands in
\figu{bands} show the results obtained with the new implementation of systematic
uncertainties, with the width of the bands illustrating the difference between
the optimistic and conservative scenarios from \Tab~\ref{tab:values}.

For the wide band beams operating at high enough energies (\lenf ,
\wbb), the old and new implementations yield very similar results. In fact, 
the bands obtained for these beams with the new implementation are
roughly included as a subset in the ones obtained with the old method,
which means that sharper predictions can be made now.  We find that
the predicted performance of \lenf\ changes only mildly, and that our results 
agree well with those presented in \Ref~\cite{Bayes:2012ex}. For \bb\
and \thk , on the other hand, there is an overall offset between the
old and new systematics treatments, and the default values (solid
curves) are not even within the old predicted ranges.  In fact, it
seems that the old treatment may have been too optimistic. As we shall
demonstrate later, the main reason for this offset is that the ratio
between $\nu_e$ and $\nu_\mu$ cross sections is needed as an external
input.  For instance, the \thk\ beam consists mainly of $\nu_\mu$, but
the $\deltacp$ measurements relies on the detection of $\nu_e$, for
which the cross sections are difficult to measure in the near
detector. The situation is precisely the opposite for \bb, for which
$\nu_e$ are produced and $\nu_\mu$ are observed at the far detector.
Note that both experiments operate at relatively low energies, where
QE and RES scattering dominate. These types of interaction have larger
uncertainties than high-energy DIS scattering, hence the large
difference between the widths of the light gray and dark colored bands
in \figu{bands} for \thk\ and \wbb. Note also that our assumption of
independent errors for the different cross section regimes introduces
an effective shape error, this being especially relevant for \bb.

The widths of the bands in \figu{bands} can also be interpreted in
terms of the robustness of the predictions.  For \lenf, the
predictions are robust with respect to systematics since all signal
channels depend on the detection of $\nu_\mu$ or $\bar \nu_\mu$, the
cross sections can be measured in the near and far detector(s).
Another interesting result is perhaps that \wbb\ outperforms \thk\ at
its default performance and is much more robust against systematic
errors. This is a result of the relatively high neutrino energies
(mostly in DIS regime) for \wbb, in combination with the wide beam
spectrum and the long baseline.

The relative position of the black lines within the new systematics
bands in \figu{bands} shows how close the default scenario is to the
optimistic performance. For example, for \wbb, the default
performance already approaches the optimistic limit, which means that
further improvements in systematics will not lead to a major increase
in sensitivity. Instead, as we will show in
\Secs~\ref{sec:performance} and~\ref{sec:impacts}, an effort towards
increasing statistics at the far detector would be more useful than a
further reduction in systematic errors. For \thk, on the other hand,
the default curve lies in the middle of the band even though it has
been simulated with exactly the same values of the systematic errors.
This means that systematics are important, and an improvement will
clearly help. We will discuss in \Sec~\ref{sec:impacts} what is the
relative importance of systematics and statistics for each of the
setups under consideration.

Comparing the no systematics limit (statistics only) with the
optimistic systematics in the new implementation (the two
uppermost/leftmost limits for each setup), one can also read off how
close the optimistic implementation is to the statistics only limit.
In almost all cases, the optimistic choice is already close to the
statistics limit, with the exception of \thk. There, even in the
optimistic case, the sensitivity to $\delta$ is limited by the
uncertainty on the QE cross section ratio.

\section{Performance comparison}
\label{sec:performance}

\begin{figure}
\begin{center}
\includegraphics[width=\textwidth]{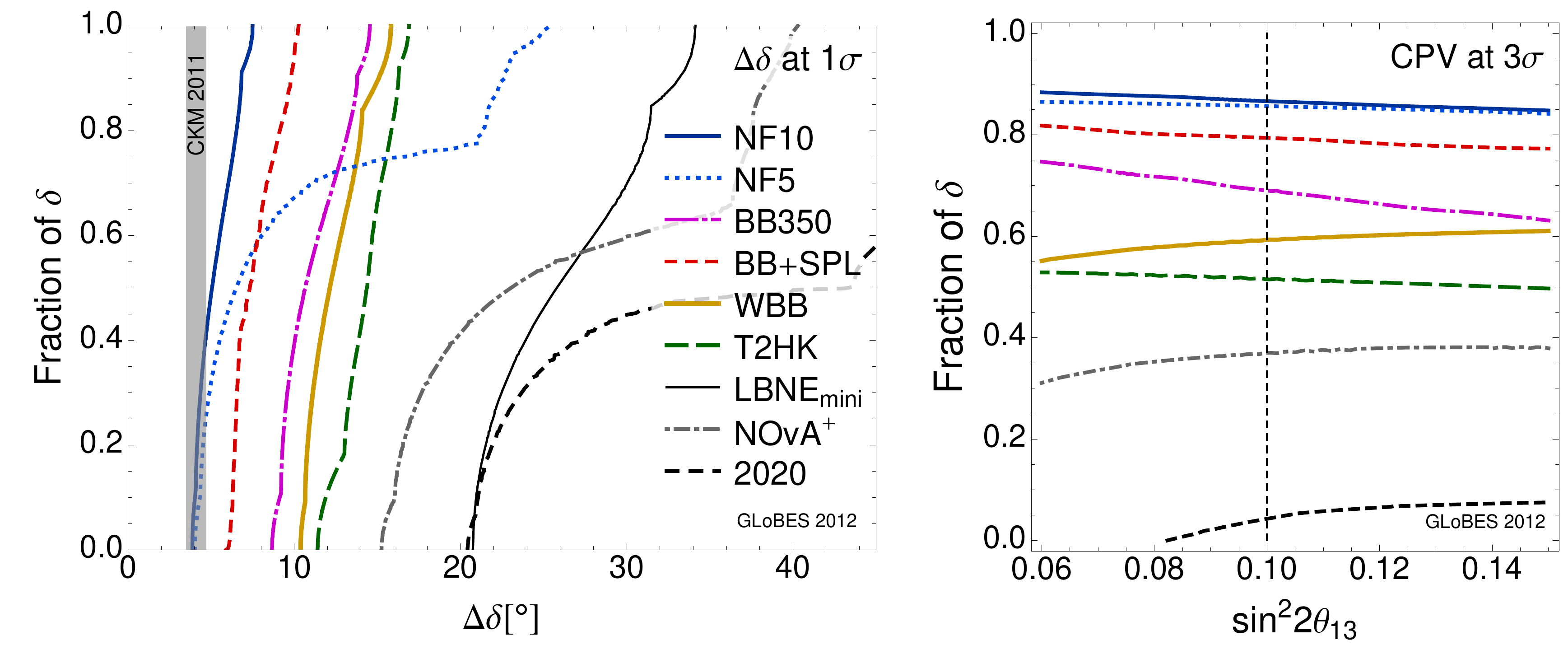}
\end{center}
\mycaption{\label{fig:defaultcpf} Comparison between the different
  setups from \Tab~\ref{tab:setups} for the default systematic errors
  listed in \Tab~\ref{tab:values} (including near detectors). We have
  also included in the comparison the results that would be obtained
  by 2020 through the combination of \tk, \nova\ and reactors. Left
  panel: Fraction of $\deltacp$ as a function of the precision at $1\sigma$ 
  for $\stheta=0.1$. Right panel: Fraction of $\deltacp$ for
  which CPV can be established at $3\sigma$ as a function of $\stheta$
  in the currently allowed range.  A true normal hierarchy has been
  assumed, and no sign degeneracies have been accounted for.  In the
  right panel, \minilbne\ is not shown because it does not reach
  $3\sigma$ sensitivity to CPV. The vertical dotted line in the right
  panel corresponds to $\stheta=0.1$, which is the true value chosen
  for $\theta_{13}$ in the left panel. In the left panel, the vertical
  gray band depicts the current precision for the CPV phase in the
  quark sector, taken from \Ref~\cite{Lenz:2012az}. }
\end{figure}

The nominal performance of all setups listed in \Tab~\ref{tab:setups}
is compared in \figu{defaultcpf}, using the default values for the
systematic errors according to \Tab~\ref{tab:values}. In the left
panel, the fraction of $\deltacp$ values for which a given precision
in $\delta$ can be achieved is shown. Considering only the benchmark
setups \bb, \lenf, \wbb, and \thk, it can be seen that the neutrino
factory outperforms all other options by a factor of two. It is the
only experiment which can achieve a precision comparable to the one
obtained in the quark sector, where the CP phase is determined to be
$\gamma=70.4^{+4.3}_{-4.4}\,^\circ$~\cite{Lenz:2012az}, depicted by the 
vertical gray band in the left panel. We also show
in this figure, in addition to the setups listed in
\Tab~\ref{tab:setups}, the results that would be obtained by the year
2020 from the combination of \tk, \nova\ and reactors.

We would like to point out the remarkable performance of \bbsb, which
outperforms any of the other superbeam and \bbeam\ options. As we
will discuss in \Sec~\ref{sec:impacts}, the reason for this is the
reduction in systematic errors related to the cross sections. For the
other alternative setups, which can be regarded as smaller versions of
the respective original proposals, the precision varies strongly as a
function of the true $\deltacp$. This is due to the fact that
intrinsic degeneracies appear around $\delta=\pm90^\circ$ superimposed
to the true solutions, effectively worsening the observable precision
on $\delta$ (see also \Ref~\cite{Coloma:2012wq}).  In the particular
case of the \lenfhs\ this is due to the coarse energy binning that we
have used for our simulations, which is fixed by the available
migration matrices.\footnote{We have checked that, if the bin size is
  reduced by a factor of two, the dependence on $\delta$ is largely
  reduced since the intrinsic degeneracies are better resolved.}

It is interesting to compare the precision on $\deltacp$ in
\figu{defaultcpf}, left panel, with the CPV discovery potential (right
panel), which mainly corresponds to the precision at specific values
of $\deltacp$ (0 and $\pi$).  Again, neutrino factories emerge as the
optimal setups, being able to observe CPV at $3\sigma$ for $\sim 85\%$
of the values of $\delta$. Compared to the left panel, \lenf\ and
\lenfhs\ perform very similar, which is expected from optimization
studies~\cite{Agarwalla:2010hk}. However, this optimization does not
include the full parameter space, which is much better covered by
\lenf\ (left panel).\footnote{In fact, the meeting point of the two
  curves in the left panel roughly corresponds to the values of
  $\deltacp$ which are relevant for a good CPV discovery potential and
  therefore the CPV sensitivity is the same.}  The
performance of almost all setups is reduced for larger $\stheta$. The
exceptions are \novaplus\ and \wbb, which benefit from the increased
statistics for larger values of $\theta_{13}$. Note that the $3\sigma$
CPV discovery potential of \thk\ is comparable to that of \novaplus ,
whereas \thk\ is clearly a better precision instrument (left panel).
As far as the comparison between \minilbne\ and \novaplus\ is
concerned, both perform similar in the left panel with \novaplus\
exhibiting a stronger dependence on $\deltacp$, as explained above.
For CPV discovery (right panel), however, \minilbne\ cannot compete at
all because it does not reach $3 \sigma$ for any value of $\delta$.
Finally, it should be noticed that the combination of \tk, \nova\ and
reactors (the 2020 line) would indeed be able to observe CPV at
$3\sigma$ for some values of $\delta$, although the fraction of
$\delta$ values for which this is possible remains well below the
$10\%$ (see also \Ref~\cite{Huber:2009cw}).

\begin{figure}
\begin{center}
\includegraphics[width=0.49\textwidth]{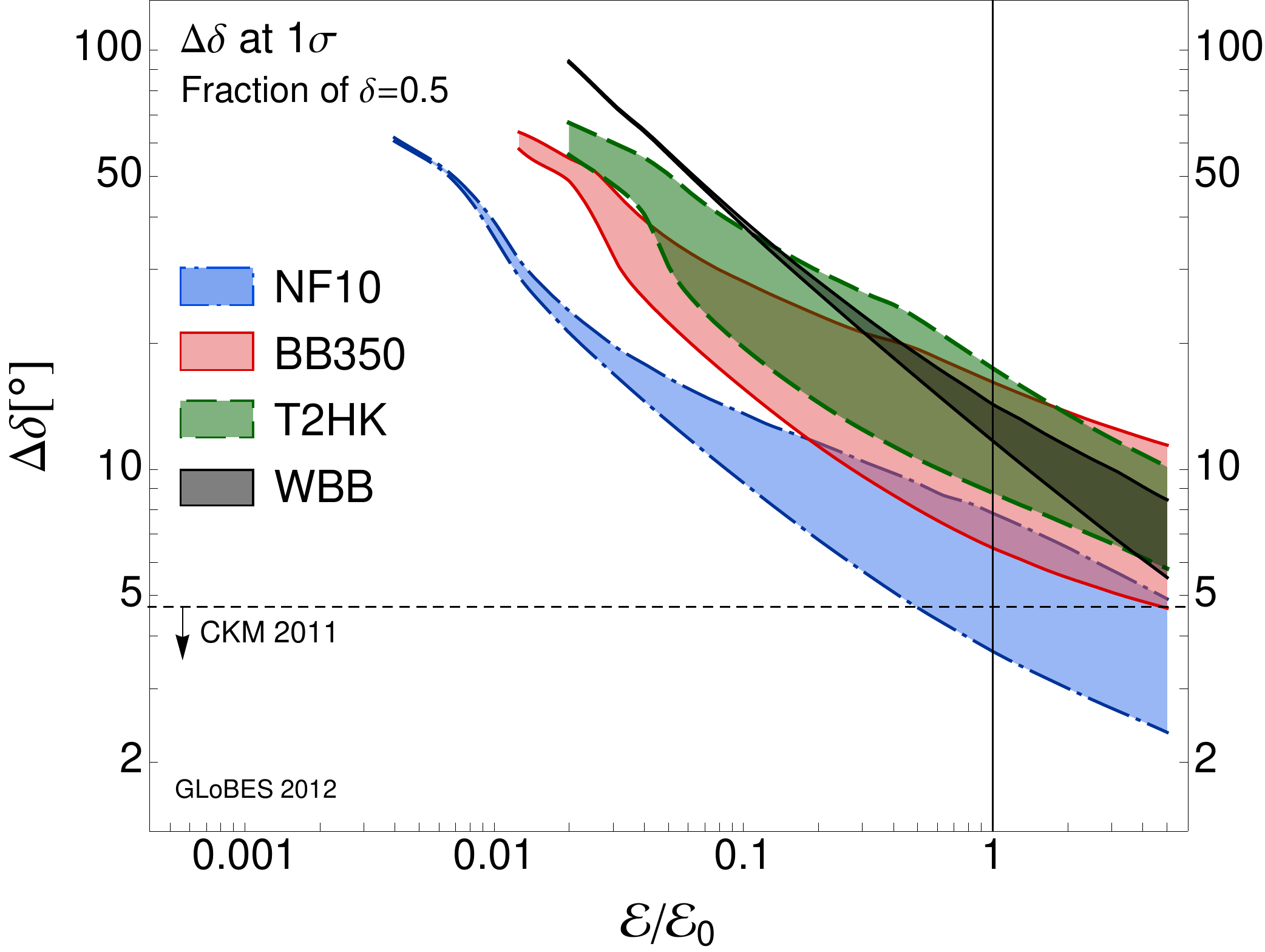} %
\includegraphics[width=0.49\textwidth]{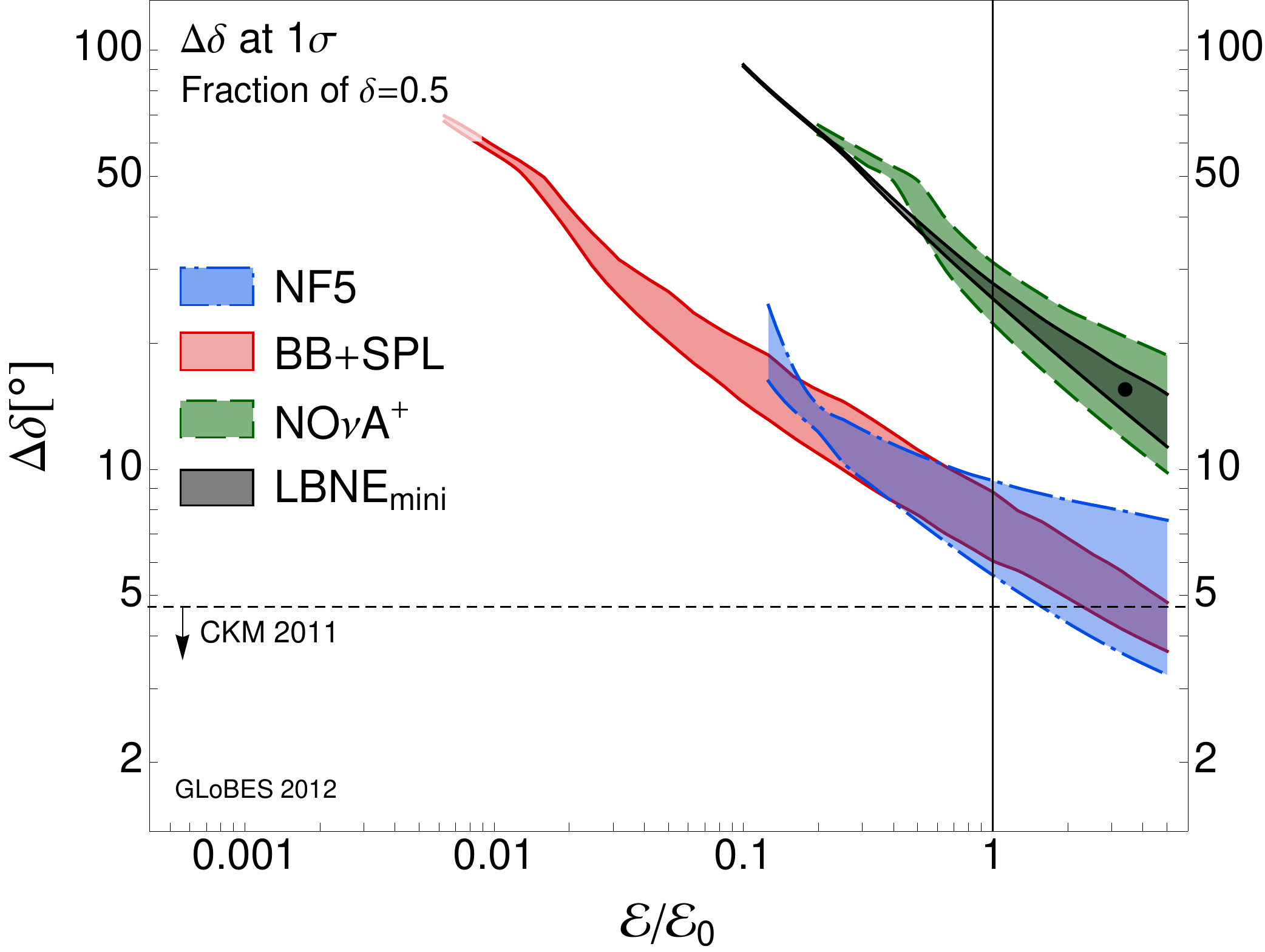}
\end{center}
\mycaption{\label{fig:lumi} Error on $\delta$ (at $1\sigma$, for $\sin^22\theta_{13}=0.1$) as a
  function of exposure, where the bands reflect the variation in the results
  due to different assumptions for the systematic errors between the optimistic
  (lower edges) and conservative (upper edges) values in
  Table~\ref{tab:values}.  In the left panel, the results for the benchmark
  setups from \Tab~\ref{tab:setups} are shown, while the right panel shows the
  results for the alternative setups. The nominal exposure $\mathcal{E}_0$, to
  which the exposure $\mathcal{E}$ on the horizontal axis is normalized, is the
  one given in \Tab~\ref{tab:setups}.  Here near detectors are included, and
  the results are shown for the median values of $\deltacp$ (fraction of
  $\deltacp$ is 50\%). The current precision on the CP phase in the CKM matrix
  $V_{\mathrm{CKM}}$ is also indicated. In the right panel, the black dot
  indicates the luminosity for the original \lbne\ configuration (34~kt
  LAr detector~\cite{Akiri:2011dv}).}
\end{figure}

The comparison between different setups does not only depend on systematic
uncertainties but also on exposure. We therefore show in \figu{lumi} the
exposure dependence of the precision on $\deltacp$ for all setups in
\Tab~\ref{tab:setups}.  Here exposure ($\mathcal{E} $) is defined as the beam
intensity (protons on target or useful parent decays) $\times$ running time
$\times$ detector mass, and only the relative exposure compared to the nominal
values $\mathcal{E}_0$ from \Tab~\ref{tab:setups} is shown. The left panel of
\figu{lumi} shows results for the benchmark setups, while the right panel
displays results for the alternative setups. The bands reflect the variation of
the performance between the optimistic and conservative choices for the
systematics uncertainties (see \Tab~\ref{tab:values}).

\figu{lumi} reveals several interesting features. First, for some
experiments the gradient at the nominal exposure is significantly
larger than for others. In particular, \wbb , \minilbne, and
\novaplus\ operate in the statistics limited regime ($\Delta \delta
\propto 1/\sqrt{\mathcal{E}}$), where the systematics contribution is
small. This makes the exposure the most relevant performance bottleneck
for them (see also \Sec~\ref{sec:impacts}). Comparing \minilbne\
with \novaplus, \novaplus\ clearly exhibits a larger dependence on
systematics. This dependence increases with exposure.  In most other cases,
when optimistic values are chosen for the systematics (lower edges of
the bands) the scaling with exposure seems to be dominated by
statistics ($\Delta \delta \propto 1/\sqrt{\mathcal{E}}$), while for
conservative values (upper edges) the setups start to be more
dominated by systematics and the curves are less steep.  In these
cases, the difference between optimistic and conservative systematics
increases significantly with exposure. An
interesting exception is \bbsb, for which systematics are equally
important regardless of the exposure. In this case, the dominant
systematics (cross section ratios) are reduced by the combination of the two
beams, so that even under conservative assumptions for the systematic
uncertainties, the performance of \bbsb\ is still dominated by statistics.

\section{Performance bottlenecks and the role of the near detectors}
\label{sec:impacts}

Here we discuss the most important limiting factors for the performance of each
individual experiment, \ie, the key factors to be watched in the design and
optimization of the experiment. As we will demonstrate, there is typically one
dominant performance bottleneck, which is however different from experiment to
experiment.  We study the impact of:
\begin{enumerate}
  \item systematic errors, including possible correlations,
  \item exposure,
  \item the near detector.
\end{enumerate}
In order to identify the key systematic errors, we start by taking all of them at
their default values and then switching off each group of systematic errors
(flux errors, cross section uncertainties, \etc) independently. 
This method reveals the systematic uncertainties that have the greatest impact, 
and which uncertainties are irrelevant for the 
measurement of $\delta$.\footnote{We have also checked that the inverse
procedure, \ie, starting from the statistics only limit and  switching on each
group of systematic errors independently, leads to  similar conclusions.
However, this second procedure is less intuitive and therefore we will not show
its results in the  following.}

\begin{figure}
  \begin{center}
    \includegraphics[width=12cm]{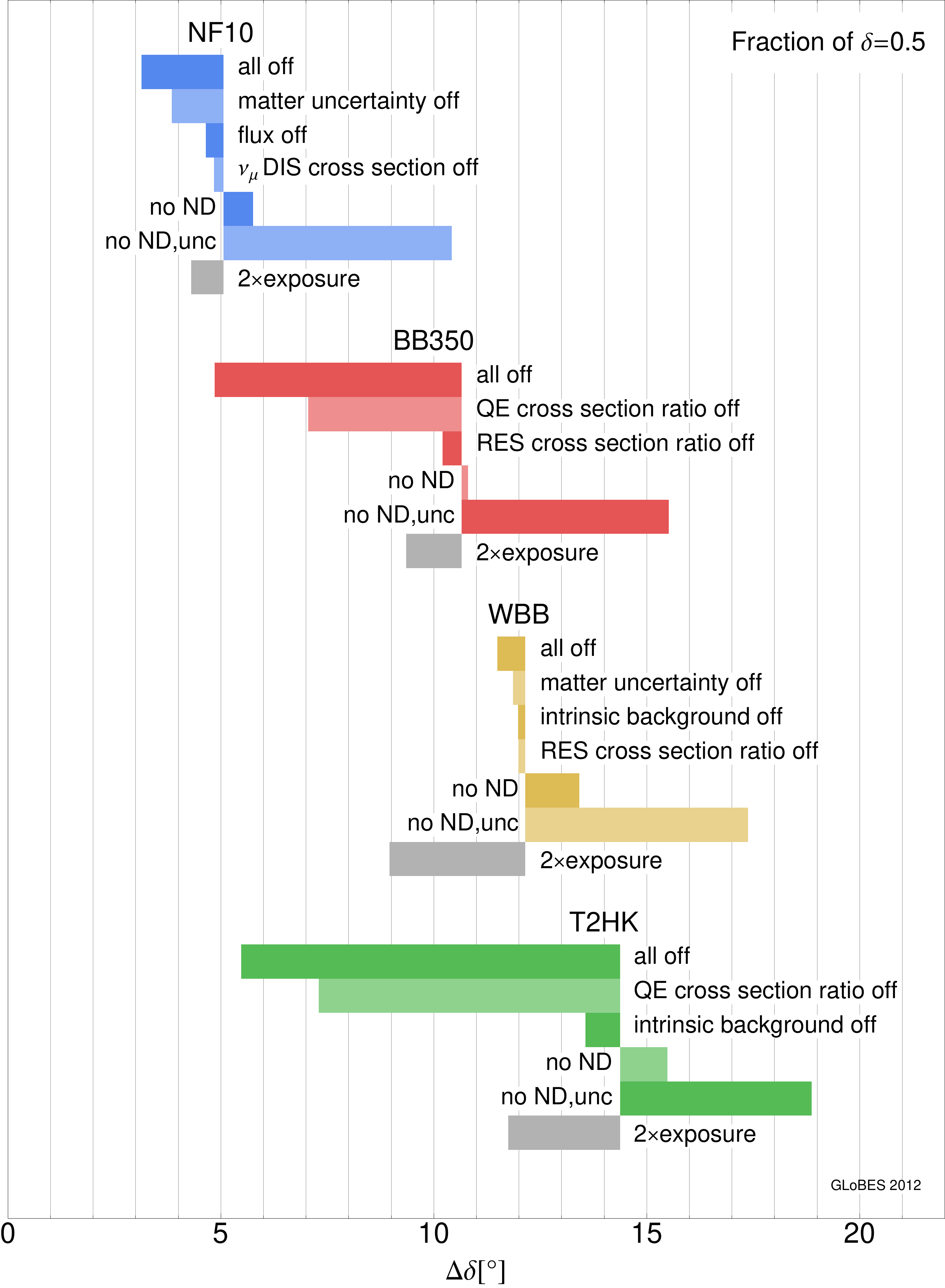}
  \end{center}
  \vspace{-0.3cm}
  \mycaption{\label{fig:bars} Dependence of the achievable precision in
    $\delta$  (at $1\sigma$, for $\stheta=0.1$) for the benchmark
    setups in \Tab~\ref{tab:setups} on systematic uncertainties, exposure, and
    near detectors. The bars show the improvement in the precision of
    $\deltacp$ compared to the default scenario if the dominant systematic
    errors are switched off separately.  Here ``all off'' refers to the
    statistics-only limit, ``matter uncertainty off'' to no matter density
    uncertainty, ``flux off'' to no flux errors, ``DIS $\nu_\mu$ cross section
    off'' to no DIS effective cross section errors for neutrinos and
    antineutrinos, ``cross section ratio off'' to fully correlated effective cross
    section errors for $\nu_e$ and $\nu_\mu$, and for
    $\bar\nu_e$ and $\bar\nu_\mu$, and ``intrinsic background off'' to no
    uncertainty on the intrinsic beam backgrounds.  The effect of doubling the
    exposure is also shown, as well as two sets of results without a near
    detector: for ``no ND'' systematic uncertainties are still correlated
    between oscillation channels
    at the far detector, while for ``no ND, unc'', also correlations between appearance
    and disappearance channels are not included.  The $\Delta\delta$ values
    shown here correspond to the median value of $\deltacp$ (\ie, for 50\% of
    $\deltacp$ values, the precision would be better, for the other 50\% it
    would be worse).
    }
\end{figure}

\begin{figure}[tp]
\begin{center}
\includegraphics[width=12cm]{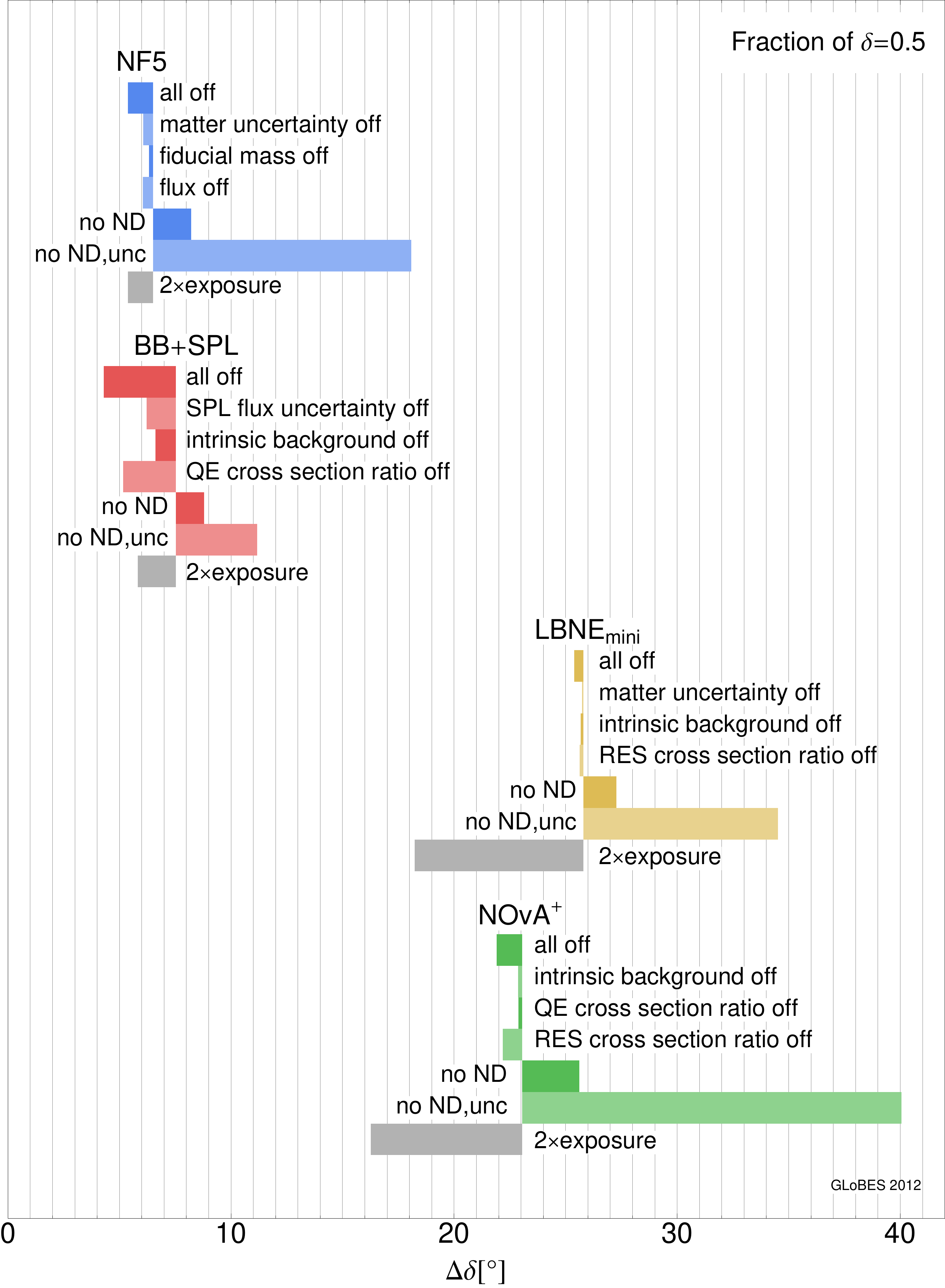}
\end{center}
\mycaption{\label{fig:bars2} Dependence of the performance of the alternative
  setups in \Tab~\ref{tab:setups} on systematic uncertainties, exposure, and
  near detectors. The meaning of the labels and abbreviations is the same as in \figu{bars}.}
\end{figure}

The impact of switching off groups of systematic uncertainties for the
different experiments is shown in \figu{bars} and \figu{bars2} for the
benchmark and alternative setups, respectively, and for the default
values of the systematics listed in \Tab~\ref{tab:values}.  In both
figures, the upper colored bars show how much the precision in
$\delta$ would improve for each experiment if a given systematic error
is switched off. (Only those groups of systematic errors that actually
have a sizeable impact for each facility are shown.) For each
experiment, the precision that would be reached in the statistics-only
limit is also shown (``all off''). As mentioned above, neither do the
different bars typically add up to the ``all off'' one, nor does the
dominant systematics alone account for it. The reason are correlations
among systematic uncertainties and between systematics and oscillation
parameters, \ie, the difference between the ``all off'' case and the
other bars can be interpreted as the importance of these correlations.
The impact of doubling the exposure (see also \figu{lumi}), as well as
the performance loss which each experiment would suffer if no near
detector was available, are also shown for each experiment.

Note that the edges of the bars shown in Figs.~\ref{fig:bars} and
\ref{fig:bars2} correspond to the medians of the corresponding
$\Delta\delta$ distributions, \ie, for 50\% of all $\delta$ values
the precision will be better than the $\Delta\delta$ value shown in
the figures, for the other 50\% it will be worse. The $\delta$ values
corresponding to the left and right edges of any given bar need not be
the same since the median may change from one edge of the bar to the
other.  Small differences in the results would appear if instead we
chose a fixed value of $\delta$ for all bars corresponding to the same
experiment, or if we sampled the $\Delta\delta$ distributions not at
their median, but at a fraction of $\delta$ other than 50\%.
Nevertheless we expect our general conclusions to remain unchanged.

Let us first discuss the impact that different systematic
uncertainties have on the benchmark setups (\figu{bars}).  For \lenf,
the most important systematic uncertainty is the one on the matter
density, as has been established earlier~\cite{Huber:2002mx}.
Improving the flux error or the understanding of the DIS cross
sections marginally helps, with the relative importance of these two
depending slightly on the value of $\deltacp$.  For \bb, the
sensitivity is limited mostly by the errors on the QE and RES effective cross
section ratios (the event rate is substantial in both regimes).
Correlations among these turn out to be important because they lead to
an effective shape error.  For \wbb, the impact of
systematics is generally small, while the sensitivity
is mainly limited by the exposure (see also \figu{lumi}). It should be noted here that 
\wbb\ has been simulated with a LAr detector, which is the least
studied among the detector technologies considered here. For instance,
it is the only detector for which no tabulated detector reponse
functions (``migration matrices'') from detailed Monte Carlo
simulations are available for the signal reconstruction up to now. For
\thk, the impact of systematic uncertainties (in particular on the QE
cross section ratio and the intrinsic beam background) is generally
large. Exposure is also important, but it is not the dominant
limitation.

An interesting question is how much the near detectors actually help
in the precision measurement of $\deltacp$. We therefore show in
\figu{bars} how the predicted $\deltacp$ precision changes when the
near detector is not included in the analysis (``no ND'').  A somewhat
surprising result is that for none of the setups considered here
omitting the near detector affects the achievable precision by more
than 1--2~degrees. Also, in none of the cases the ND is the most
critical factor. The main reason is that, even without a near
detector, most systematic errors are correlated among the different
oscillation channels, so that the nuisance parameters are constrained
by the requirement of self-consistency among the different far
detector channels (in particular appearance and disappearance).  Note
that this self-consistency requirement relies entirely on the validity
of the three active flavor oscillation framework.
Thus, in the absence of a near detector it is doubtful that meaningful bounds on
physics beyond the Standard Model (such as sterile neutrinos or non-standard 
interactions~\cite{Kopp:2007ne,Antusch:2009pm}) could be obtained. 

In order to illustrate the importance of correlations between
appearance and disappearance data, we also show in \figu{bars} results
for the case where the near detector is omitted and in addition
correlations between appearance and disappearance channels in the far detector are not
included, \ie, the appearance and disappearance data sets are assigned
independent systematic errors (``no ND, unc'').  In this case, as
expected, the near detector plays a crucial role for all setups. The
difference between these bars and the bars labeled ``no ND'' shows
explicitly the importance of correlations and how disappearance data
can be used to constrain systematic errors in the appearance sample in
a very efficient way. The effect of the disappearance channel is
particularly relevant for the \bb\ setup, for instance. In this case,
since $\nu_e$ disappearance is small, the far detector is particularly
useful in constraining systematic effects related to flux
uncertainties and $\nu_e$ cross section measurements.  Therefore, the
near detector does not provide any additional information and can be
removed from the analysis with practically no impact on the precision.
In addition, for \bb\ (\thk) the near detectors do not provide a
measurement of the $\parenbar\nu_\mu$ ($\parenbar\nu_e$) cross
sections needed for the far detector appearance measurement, but only
of the flavor-conjugate cross section.  (In \thk, the near detector is
in principle sensitive also to the $\parenbar{\nu}_e$ cross sections
due to the the intrinsic beam backgrounds, but statistics in these
channels is too small to allow for a precise cross section
measurement.) The experiment which benefits most from the near
detector is \wbb, where having the near detector is in principle more
important than improving any of the systematic errors. The reason for
this is that this setup is statistically limited, and is therefore not
able to constrain both the nuisance parameters and the atmospheric
oscillation parameters independently in the analysis. Thus, increasing
the nominal exposure would be much more critical in this case.

\figu{bars2} shows how the performance of the alternative setups in
\Tab~\ref{tab:setups} depends on systematics, exposure, and the near
detectors.  We notice that for \lenfhs , the relative impact of the
systematic errors (including the matter density uncertainty) is
smaller whereas exposure is somewhat more important than for \lenf. In
addition, \lenfhs\ is the only experiment for which the near detector
is more important than systematics or exposure. For \minilbne\ and
\novaplus, the near detector has also a larger effect than the systematic
uncertainties, but the main limitation for these statistics-dominated
experiments is exposure. In fact, an increase in statistics may also
render the near detector unnecessary because, similar to \wbb\
discussed above, \minilbne\ and \novaplus\ need the near detector
mainly because they are unable to constrain both the systematic
nuisance parameters and the atmospheric oscillation parameters
independently using disappearance data.  Thus, in the optimization of
experiments of this type, the benefits of a near detector and of
increased statistics have to be carefully weighed against each other.

An interesting question one could ask at this point is whether there
are feasible ways of reducing systematic uncertainties, especially the
ones on the QE cross sections. An interesting example for this is
\bbsb. In this setup, both $\nu_e$ and $\nu_\mu$ cross sections can
be measured precisely in the same detector, which reduces the impact
of systematics and increases the absolute performance. This can be
seen from the reduced length of the ``all off'' bar for \bbsb\
compared to the \bb\ or \thk\ cases, which all operate in the low
energy regime.  Some further improvement would be achieved if the \spl\
flux uncertainty was reduced, though. Note that \bbsb\ could in
principle even compete with \lenfhs\ if the exposure could be
significantly increased, the cross section ratios could be better
constrained, or the \spl\ flux could be better understood.  This shows
how a combination of facilities can be of great help in reducing the
impact of systematics on their performance. A similar effect would be
obtained if an independent measurement of the $\nu_e/\nu_\mu$ cross
section ratio was performed for both neutrinos and antineutrinos. The
proposed low energy muon storage ring experiment
$\nu$-STORM~\cite{nustorm} would be ideal for this measurement. 

An additional method to reduce the impact of systematics could be a facility
optimized for the second oscillation peak, see for instance
\Ref~\cite{Coloma:2011pg}, where an \spl-based experiment with a
detector at 650~km instead of 130~km is proposed. This would be useful
to increase the CPV discovery potential of the facility as well as to
reduce the impact of systematic errors. Note that in
\Ref~\cite{Coloma:2011pg}, correlations between systematic
uncertainties were not taken into account and near detectors were not
simulated explicitly. We have checked that the conclusions still hold
in the case where full correlations are taken into account.  Indeed,
the setup proposed in~\cite{Coloma:2011pg} exhibits the least
dependence on systematic errors among all the experiments compared
here.

\section{Summary and conclusions}
\label{sec:summary}

Systematic uncertainties in neutrino oscillation experiments are
especially important for large $\theta_{13}$. Hence, a dedicated
comparison with a careful treatment of these uncertainties is needed
to determine the optimal next-generation experiment, given the large
value of $\theta_{13}$. Also, the degree to which this optimization
depends on the assumptions regarding systematic uncertainties should
be carefully assessed.

In this study, we have analyzed and compared superbeams, \bbeams,
and neutrino factories on an equal footing, paying special attention
to systematic uncertainties.  In particular, a realistic
implementation of systematic uncertainties in the simulations used to
predict the sensitivity of future experiments depends not only on
individual numbers for certain systematic errors, but also on how
these errors are correlated among different detectors, oscillation
channels, \etc\ In most previous studies, only few types of systematic
uncertainties were considered, and the respective error margins were
chosen in order to account in an effective way for the real error
menu. In particular, near detectors were typically not simulated
explicitly, and correlations were neglected. In this paper, instead,
we have used explicit near and far detector simulations with
comparable assumptions and an improved systematics implementation
which takes into account all possible correlations.  Moreover, to
allow for a simple and fair comparison of different facilities, we
have used identical assumptions on external input (in particular cross
sections) wherever possible. Besides our default set of systematic
errors, we also consider more conservative and a more optimistic
scenarios (see \Tab~\ref{tab:values}), which should encompass the
performance of a real experiment.

\Tab~\ref{tab:setups} summarizes the setups studied in this work. 
Since we expect that the mass hierarchy can de determined by all of
the discussed setups (with the possible exception of \novaplus), we
have used the $3\sigma$ discovery potential for leptonic CP violation (CPV)
and the achievable precision at $1\sigma$ in the measurement of $\delta$ ($\Delta\delta$)
as our main performance indicators.  While the first indicator depends
on the performance of the experiment around the specific values
$\deltacp = 0$, $\pi$, the second one treats all values of $\deltacp$
as equally important.  Since the dependence of the experimental
sensitivity on $\deltacp$ is in general complicated, we present our
results in terms of the fraction of $\deltacp$ values for which a
certain precision (or better) is achieved, see \figu{cpfraction} for
illustration.

We have compared our new systematics implementation with the previous
effective treatment and have found good agreement except for \thk\ and
\bb, for which the near--far extrapolation depends strongly on the
poorly known ratio of $\nu_e$ and $\nu_\mu$ QE cross sections.
Therefore, the performance of these experiments strongly depends on
the systematics assumptions, and it is difficult to make
self-consistent predictions. We have also discussed the impact of the
true value of $\deltacp$ on the measurement. While the performance is
relatively uniform for most experiments, especially the precision
attainable from \novaplus\ and \lenfhs\ depends on $\deltacp$.
\lenfhs, for instance, was clearly optimized for CPV, whereas
a precise measurement independent of $\deltacp$ would require higher
muon energies and longer baselines, as realized in the \lenf\ setup.
Interestingly, \thk\ does not exhibit such a strong dependence on
$\deltacp$, in spite of the narrow beam spectrum.

For each experiment under consideration here, we have also identified
the main limitations to a further increase in sensitivity.  We have
considered systematic uncertainties, exposure, and the impact of the
near detector as possible bottlenecks. The results can be summarized
as follows:
\begin{description}
\item[Superbeams] can be divided into two classes: low and high
  energy. For the low-energy experiment \thk, the fact that the
  $\nu_e$ cross sections needed for appearance measurements cannot be
  easily obtained from the near detector, is clearly the most
  important limitation. This is especially relevant since it operates in
  the QE regime, where cross section uncertainties are large and it is
  very difficult to relate the measured $\parenbar\nu_\mu$ cross
  sections to the $\parenbar\nu_e$ cross sections needed for the
  appearance measurement.  Although the intrinsic beam backgrounds
  were included in our near detector simulations, we could not
  identify a simple way of measuring the $\parenbar\nu_e$ cross
  sections directly with the required precision.  Uncertainties in the
  the intrinsic beam background and the limited exposure are also
  limiting factors for \thk, and we find that the availability of a
  near detector is of some importance.  For \wbb , \minilbne , and
  \novaplus , which operate at higher energies, systematic errors can
  be controlled to the level needed, which in turn implies that, from
  the systematics point of view, very robust predictions can be made
  for these experiments.  The critical issue for them is instead
  exposure. For instance, for \minilbne, investing in the far detector
  mass may be more important than constructing a near detector. We
  conclude that for superbeams, the impact of systematic uncertainties
  depends mainly on the beam energy, especially because cross section
  uncertainties are much smaller in the high energy DIS region than in
  the low-energy QE region. The separation into narrow-band beams
  (\thk , \novaplus ) and wide-band beams (\wbb , \minilbne ), on the
  other hand, has turned out not to be the primary issue.

\item[Beta-beams] using $^6$He and $^{18}$Ne for the neutrino production also
  suffer from the fact that the ratio between the $\parenbar\nu_e$ and
  $\parenbar\nu_\mu$ QE cross sections is needed as an external input.  If a
  low-$\gamma$ \bbeam\ is however combined with an \spl-based superbeam
  (\bbsb), the performance is much better and much more robust than the one of
  a high-$\gamma$ \bb. In fact, \bbsb\ is the only experiment that could
  compete with a neutrino factory. The reason is that \bbsb\ uses both the
  $\parenbar\nu_e \rightarrow \parenbar\nu_\mu$ and $\parenbar\nu_\mu
  \rightarrow \parenbar\nu_e$ channels, so that both $\parenbar\nu_e$ and
  $\parenbar\nu_\mu$ cross sections can be measured.

\item[Neutrino factories] achieve the best absolute precision (comparable to that in the quark sector) and are very
  robust with respect to systematic errors. This is due to two main factors: firstly, the energy of the beam lies in the DIS regime where cross section uncertainties are small; secondly, this is the only experiment where the final flavor cross section can be determined in a self-consistent way from the disappearance data.
  Depending on baseline and muon energy,
  the most relevant factors affecting their performance are the matter density
  uncertainty (for setups with longer baselines and high $E_\mu$, such as
  \lenf) or exposure and near detector (for setups with shorter baselines and
  low $E_\mu$, such as \lenfhs).
\end{description}

We also find, remarkably, that near detectors have a relatively small
impact and help to improve the precision in $\delta$ by only about
1--2$^\circ$ or less. The reason is that most systematic uncertainties
are correlated between appearance and disappearance channels and can
therefore be constrained by the far detector alone, provided that
statistics in the disappearance channel is good enough to break
correlations between systematic effects and the atmospheric
oscillation parameters.  The near detector turns out to be practically
useless at \bbeam\ facilities: since the $\nu_e$ disappearance
channel does not depend very much on the atmospheric parameters, the
$\nu_e$ data from the far detector are even more useful for
constraining flux and cross section uncertainties than in other
experiments. It should be kept in mind, however, that near detectors
will still be required to constrain effects of new physics in neutrino
oscillations.  
In addition, if a combined analysis of appearance and disappearance data is not
possible, a near detector proves to be critical in order to constrain
cross section and flux uncertainties, as expected. Moreover, a well
designed near detector facility is an excellent safeguard against
``unknown unknowns''. 

The most attractive superbeam option, as far as the impact of systematic uncertainties is concerned, is a high-energy wide-band beam like \lbno\ or \lbne\ operating in the DIS regime. The \minilbne\ experiment may be
the first step towards such an experiment. However, both \lbno\ and \lbne\
have a limited discovery potential for CPV, and would suffer
strongly from a reduction in statistics (notice that the \wbb\ setup
studied here had a LAr detector with a fiducial mass of 100~kton). The
ultimate precision can be reached with a neutrino factory, which is
the only experiment with a precision competitive to the one achieved
in the quark sector. The \bbsb\ combination of a $\gamma=100$ beta
beam and a superbeam is a very interesting option that is very robust
with respect to systematic errors and has a performance closer to
neutrino factory than any other superbeam or \bbeam. Previous
studies have underestimated the performance of \bbsb\ because they did
not take into account correlations between systematic uncertainties.
Predictions for \thk\ and \bb\ heavily rely on external input on the
flavor dependence of the QE cross sections.  Here, an independent
$\nu_e$ cross section measurement, for instance by a facility like
$\nu$-STORM, is necessary.

\subsubsection*{Acknowledgments}

We would like to thank E.~Fernandez-Martinez for illuminating
discussions, and M.~Mezzetto who contributed during the early stages
of this work. We thank M.~Vagins for providing the \thk\ fluxes in
machine readable format. WW would like to acknowledge support from DFG
grants WI 2639/3-1 and WI 2639/4-1.  This work has been supported by
the U.S.  Department of Energy under award number
\protect{DE-SC0003915}. PC, PH and WW would like to thank GGI Florence
for hospitality during their stay within the ``What's $\nu$?''
program. PC would also like to thank CERN and Fermilab for their
hospitality during completion of this work.  This work has been also
supported by the EU FP7 projects EURONU (CE212372) and INVISIBLES
(Marie Curie Actions, PITN-GA-2011-289442).  Fermilab is operated by
Fermi Research Alliance under contract DE-AC02-07CH11359 with the
United States Department of Energy.

\appendix

\section{Simulation details for experimental setups}
\label{sec:sim}

This appendix summarizes the technical details of our simulations for each of
the setups included in our study.

The main parameters of all setups, \ie, baselines, detector technology,
fiducial masses, \etc, are summarized in \Tab~\ref{tab:setups}, and the
corresponding references are given in \Sec~\ref{sec:setups}.  The
beam powers quoted in \Tab~\ref{tab:setups} are based on the cited references,
but note that in some cases they were computed from an anticipated running
time and a number of protons on target (pot). This requires an assumption
on the experimental duty cycle, \ie, the number of useful seconds per year:
\thk\ assumes 130
useful days per year ($1.12\times10^7$ secs, approx.); \novaplus\ assumes
1.7$\times 10^7$ sec$\times$yr$^{-1}$; \minilbne\ assumes $2\times 10^7$
sec$\times$yr$^{-1}$; and \bbsb\ and \wbb\ assume $10^7$ sec$\times$yr$^{-1}$.
The maximum running time
has been restricted to ten years for all experiments. This is usually split
into neutrino and antineutrino running, except for neutrino factories where
both muon polarities are circulating inside the decay ring at the same time.

The setups considered in this paper were defined as close as possible to the
existing ones in the literature, \ie, no further optimization with respect to
beam or detector parameters (efficiencies, energy resolutions, \etc) was done.
Such a study would be especially relevant for the setups with LAr detectors,
since their performance is still uncertain. We have simulated the LAr detector
performance flat signal efficiencies and NC rejection efficiencies from
Refs.~\cite{Coloma:2012ma,Coloma:2012ut} for \wbb\, and from
\Ref~\cite{Akiri:2011dv} for \minilbne\ and \novaplus. In all cases, the same
energy resolution as in \Ref~\cite{Akiri:2011dv} has been considered, and NC
backgrounds have been migrated to low energies using matrices from the \lbne\
collaboration~\cite{Akiri:2011dv}. In the absence of any information regarding
the performance of a LAr detector at the surface, we have it to be the same as
for an underground LAr detector.  We have used migration matrices (tabulated
detector response functions) to simulate the detector performance for neutrino
factories~\cite{Laing:2010zza}, \bbeams~\cite{BurguetCastell:2005pa} and the
\spl~\cite{Campagne:2006yx}. The signal reconstruction for \thk\ has been
simulated following \Ref~\cite{Huber:2007em}, while the flux has been taken
from \Ref~\cite{Abe:2011ts}.

The following sources of backgrounds have been considered in our analysis. For
neutrino factory-based setups, only NC and charge mis-identification
backgrounds have been included, since these are the only relevant backgrounds
according to recent simulations of the MIND~\cite{Laing:2010zza,Bayes:2012ex}. For
conventional beams and superbeams, we have followed the references mentioned in
\Sec~\ref{sec:setups} and have included as main backgrounds NC events
mis-identified as CC events and intrinsic contamination of the beam. Finally, only NC
backgrounds have been considered for \bbeams. It is well-known that
atmospheric neutrinos could also constitute a relevant source of background for
this kind of experiments because suppression factors below $10^{-3}$ are
difficult to achieve. This is especially true for the low-energy \bbeam\ at
$\gamma=100$. We have nevertheless omitted the atmospheric background in
our simulations since \Ref~\cite{FernandezMartinez:2009hb} showed that it is
only relevant for small $\theta_{13}$, while for large values of $\theta_{13}$,
suppression factors as large as $10^{-2}$ can be tolerated.

For high energy $\nu_\mu$ beams, there is an additional background coming from
the production of $\tau$ leptons in the far detector. All facilities considered
in this paper are located around the first atmospheric oscillation maximum
where $\nu_\mu \rightarrow \nu_\tau$ oscillations are strong. For high energy
beams such as the \lenf\ or \wbb, the $\nu_\tau$ energy is sufficient to
produce a considerable amount of $\tau$ leptons at the detector.  The leptonic
$\tau$ decays $\tau \to e\nu\nu$ and $\tau \to \mu\nu\nu$, which have a
branching ratio of 17\% each, can lead to events that mimic $\nu_\mu$ or
$\nu_e$ charged current interactions.  This phenomenon, known as the
$\tau$-contamination, has been studied in the context of the high-energy
(25~GeV) neutrino factory~\cite{Indumathi:2009hg,Donini:2010xk,Dutta:2011mc},
for non-magnetized detectors~\cite{Tang:2011wn}, and for wide band
beams~\cite{LBNOeoi}.  It was shown in \Ref~\cite{Agarwalla:2010hk} that $\tau$
contamination does not constitute a problem for the golden channel at a
neutrino factory, and \Ref~\cite{Indumathi:2009hg} shows that in the disappearance
channel, it would only be problematic for precision measurements in the
atmospheric sector.\footnote{If the impact in the atmospheric sector is very
large, it may have an indirect effect in the achievable precision in $\delta$.}

Contamination by $\tau \to e\nu\nu$ decays could in principle also affect high
energy conventional neutrino beams. In the case of a typical
superbeam such as \wbb, with its flux peaking around 4-5~GeV, the majority of
the fake events would be reconstructed with an energy below 1.5--2~GeV and
would therefore affect mainly the second oscillation maximum. This could affect
the measurement of $\delta$. However, it was shown in \Ref~\cite{Huber:2010dx}
that for a wide-band beam with a detector located at the first oscillation
maximum, the second maximum is of little importance anyway because it is
strongly affected by NC backgrounds.  Note also that $\nu_\tau$ cross sections
have large uncertainties, which could contribute significantly to the overall
systematic error budget.  Kinematic cuts on the momentum distributions of the
visible final state particles, which might help to reduce the $\nu_\tau$
contamination in superbeam experiments, are currently under
investigation~\cite{LBNOeoi}.

From these arguments it is clear that a dedicated study is required to
address the actual impact of $\tau$ contamination on
precision measurements, both at neutrino factories and at conventional
neutrino beams. However, since this is beyond of the scope of the
present paper, we have not included $\tau$-related
backgrounds for any of the setups studied in this work.

\Tab~\ref{tab:events} shows the total number of events expected at the
far detectors of the setups under study in this paper. In each case, the size
of the near detector
has been chosen such that the results are
dominated by the statistics at the far detector. For this purpose, we
require at least 10 times more disappearance events at the near
detector compared to those obtained at the far detector.  In the case
of very long baseline setups with high density detectors (for
instance, \lenf\ or \wbb ) 25~tons would be enough to fulfill this
requirement. However, for setups with short baselines and very massive
far detectors (\thk, \bb, \bbsb) the near detector mass had to
be increased to 50, 100 and even 1\,000 tons.

\begin{table}
\begin{center}{
\renewcommand{\arraystretch}{1.5}
\begin{tabular}{cl|cccc| }
& Setups & $\nu$ app & $\bar\nu$ app & $\nu$ dis & $\bar\nu$ dis \\ \hline

\multirow{4}{*}{\begin{sideways} Benchmark \phantom{a}  \end{sideways}} &
 \lenf\	 	  &  44880/35 & 8701/61 & 159532/19 & 209577/21 \\
&\bb\ 		  & 2447/378 & 2262/330 & 93775/-- & 106750/-- \\
&\thk\	 	  & 4754/2106 & 2006/2290 & 33788/544 & 168685/5502 \\
&\wbb\	 	  & 1830/248 & 147/148 & 5526/763 & 1884/515 \\ \hline

\multirow{4}{*}{\begin{sideways} Alternative \phantom{lala}  \end{sideways}}  &
\lenfhs\	  & 11022/4 & 2916/11 & 18337/2 & 32891/2 \\
&\bblow\ 	  & 1203/96 & 1048/81 & 65926/-- & 44776/-- \\
& \spl\		  & 10455/1546 & 4453/1695 & 214524/9 & 93039/4 \\
&\minilbne\ & 389/162 & 63/102 & 3330/533 & 941/1419 \\
&\novaplus\ &  752/590 & 155/386 & 7335/1255 & 3179/2397 \\
\hline
\end{tabular}}
\end{center}
\mycaption{\label{tab:events} Total number of signal/background events expected
  at the far detectors of the experiments considered here. Numbers have been obtained
  assuming $\theta_{12}=32^\circ$, $\theta_{23}=45^\circ$, $\theta_{13}=9^\circ$,
  $\delta=0$, $\Delta m_{21}^2=7\times 10^{-5}$ eV$^2$ and $\Delta
  m_{31}^2=3\times 10^{-3}$ (normal hierarchy). Disappearance channels at \bbeams\ 
  have been assumed to be background-free.}
\end{table}

As mentioned in \Sec~\ref{sec:setups}, we assume that the near detector is
located sufficiently far away from the neutrino source for the neutrino spectra
at the near and far sites to be similar.  In the context of a neutrino factory,
\Ref~\cite{Tang:2009na} demonstrates that a detector at a distance of 1~km from
the end of a 600~m long decay straight can be simulated with an effective
baseline (relative to an imaginary pointlike source) of 1.27~km. In this case,
geometry effects arising from the straight section of the decay ring and the
detector extension are expected to be small. We have therefore adopted this
value for the neutrino factory-based setups. Similar values have been
considered for superbeam setups since the decay pipe for pions would be
comparable in size to the storage ring of a neutrino factory, or even smaller.
In the case of \bbeams\, on the other hand, the storage ring would need
straight sections around 2\,500~m in length to keep its livetime (\ie, the
useful fraction of the decay ring, $\ell =\frac{L_s}{L_t} $, where $L_s$ and
$L_t$ are the lenght of the straight sections and the total lenght of the decay
ring, respectively) around 35\% (see, for instance,
\Ref~\cite{Choubey:2009ks}). Therefore, we take the effective near detector
baseline to be 2~km.

Finally, we have assumed the near and far detectors to be identical regarding
signal and background rejection efficiencies as well as bin sizes and energy
resolution. However, this may not apply in general, and some
differences may arise as a consequence of a different detector technology
or design. In particular, a different detector technology may imply a
different background rejection efficiency. In all cases under study, the most
relevant source of background are NC events misidentified as CC events.
Therefore, we have (conservatively) assumed that the systematic uncertainties
on this background are uncorrelated between the two detectors. However, for the
sake of simplicity we have assumed the rejection efficiencies for this
background to be the same for the two detectors in all cases. The rest of
systematic uncertainties have been taken to be fully correlated between near
and far detectors (with the exception of those affecting fiducial masses,
obviously).

\section{Details on systematics implementation}
\label{sec:sysdet}

The standard implementation of systematic uncertainties in
GLoBES~\cite{Huber:2004ka, Huber:2007ji, Huber:2006vr}, based on the pull
method~\cite{Huber:2002mx, Fogli:2002pt}, has been extended for this work to
allow for an easier and more realistic treatment. For each GLoBES
rule $r$,\footnote{A rule corresponds to a realistic data set including typically
several experimentally indistinguishable signal and background components
(``channels''). For example, in a superbeam using a WC detector, rules could
correspond to electron-like and muon-like events.  The channels contributing to
each of these would would be actual $\nu_e$ and $\nu_\mu$ interactions on the
one hand, and background from beam contamination, neutral current interactions,
flavor mis-identification, \etc , on the other hand.} we define a
Poissonian $\chi^2$ according to
\begin{equation}
  \chi_r^2 = \sum_i  2 \bigg( T_{r,i}(\vec{\Theta}, \vec{\xi}) - O_{r,i}
                            + O_{r,i} \ln \frac{O_{r,i}}{T_{r,i}(\vec{\Theta}, \vec{\xi}))} \bigg) \,.
\label{equ:chirule}
\end{equation}
Here $T_{r,i}(\vec{\Theta}, \vec{\xi})$ is the predicted number of
events in the $i$-th energy bin for this rule, for a particular set of
oscillation parameters $\vec{\Theta}$ and systematic biases
(``nuisance parameters'') $\vec{\xi}$.  $O_{r,i}$ is the ``observed''
event rate, \ie\ the rate corresponding to the set of assumed ``true''
oscillation parameters.  Both $T_{r,i}$ and $O_{r,i}$ receive
contributions from different oscillation channels $c$:
\begin{equation}
  T_{r,i}(\vec{\Theta}, \vec{\xi}) = \sum_c \big(1 + a_{r,c}(\vec{\xi}) \big) S_{r,c,i}(\vec{\Theta}) \,.
\end{equation}
Here, $S_{r,c,i}(\vec{\Theta})$ is the event rate that channel $c$
contributes to rule $r$.  The rule- and channel-dependent auxiliary
parameters $a_{r,c}$ are given by
\begin{equation}
  a_{r,c} \equiv \sum_k w_{r,c,k} \, \xi_k \,,
\end{equation}
where the coefficients $w_{r,c,k}$ (which can be either zero or one)
specify if a particular nuisance parameter $\xi_k$ affects the
contribution from channel $c$ to rule $r$ ($w_{r,c,k}=1$) or not
($w_{r,c,k}=0$). Thus the $w_{r,c,k}$ determine how systematic
uncertainties are correlated among the rules and channels, and thus
among detectors, beam polarities, flavors, signal and background, 
\etc\ For instance, in the case of a fiducial mass error, one would choose
$w_{r,c,k}=1$ for all rules describing data from a particular
detector, and $w_{r,c,k}=0$ for all other rules.  In the fit, the
nuisance parameters $\xi_k$ are minimized over along with the
oscillation parameters, and so-called \textit{pull} terms are added to
the $\chi^2$ to ensure that their magnitude cannot get much larger
than the systematic uncertainties $\sigma_k$ they are implementing.
The total $\chi^2$ is
\begin{equation}
  \chi^2 = \sum_r \chi_r^2 + \sum_k  \bigg( \frac{\xi_k}{\sigma_k} \bigg)^2 . \, 
\end{equation}
The important new feature compared to the standard treatment of
systematic uncertainties in GLoBES is that the nuisance parameters
$\xi_k$ are no longer associated with a particular rule, but are
defined globally and can in principle affect any rule or channel.


\providecommand{\href}[2]{#2}\begingroup\raggedright\endgroup

\end{document}